\documentclass[aps,prd,twocolumn,showpacs,superscriptaddress,nofootinbib]{revtex4-2}

\usepackage{graphicx}
\usepackage{dcolumn}
\usepackage{bm}

\usepackage{overpic}
\usepackage[english]{babel}
\usepackage{graphicx}
\usepackage{epsfig}
\usepackage{amssymb}
\usepackage{amsmath}
\usepackage[plainpages=false,pagebackref=false,pdftex]{hyperref}
\usepackage{placeins}
\hypersetup{colorlinks=true, linkcolor=blue, citecolor=blue,urlcolor=blue}
\usepackage{ulem}
\usepackage[T1]{fontenc} 
\usepackage[utf8]{inputenc}
\usepackage{color}

\usepackage[titletoc]{appendix}

\begin{document}

\preprint{APS}

\title{Pion quasiparticles in isospin medium from holography}

\author{Weijian Liang}
\author{Xuanmin Cao}%
 \email{Corresponding author: caoxm@jnu.edu.cn}
 \author{Hui Liu}%
 \author{Danning Li}%
 \email{Corresponding author: lidanning@jnu.edu.cn}
\affiliation{%
Department of Physics and Siyuan Laboratory, Jinan University, Guangzhou 510632, China
}%





\begin{abstract}
The properties of the pion quasiparticle in hot and dense isospin medium, including the screening mass, pole mass and thermal width, as well as their relationships with the pion superfluid phase transition, are investigated in the framework of two-flavor ($N_{f}=2$) soft-wall AdS/QCD models. We extract the screening mass of the pion from the pole of the spatial two-point Retarded correlation function. The screening masses of both neutral and charged pions increase monotonously with the increasing of temperature. However, the isospin chemical potential $\mu_{I}$ would depress the screening masses of the charged pions, $m_{\pi^{\pm},\rm{scr}}$. With the increasing of $\mu_{I}$, $m_{\pi^{\pm},\rm{scr}}$ monotonically decrease to zero on the boundary between the normal phase and the pion superfluid phase, while the screening mass of the neutral pion, $m_{\pi^0,\rm{scr}}$, remains almost unchanged. We also extracted the pole mass $m_{\rm{pole}}$ and thermal width $\Gamma$ of the pion from the pole of temporal two-point Retarded correlation function, i.e., the corresponding quasi-normal frequencies, $\omega=m_{\rm{pole}}-i\Gamma/2$. The pole masses of the three modes ($\pi^0, \pi^+, \pi^-$) are splitting at finite $\mu_{I}$. The thermal widths of the three modes increase with temperature. Furthermore, the pole mass and thermal width of $\pi^+$ decreases almost monotonically with the increasing of $\mu_{I}$, reaching zero at $\mu_{I}=\mu_{I}^c$, simultaneously. It indicate that $\pi^+$ becomes a massless Goldstone boson as a result of the pion superfluid phase transition.  
\end{abstract}

\maketitle


\section{\label{sec:level1}Introduction}

The fundamental theory of the strong interaction is Quantum Chromodynamics (QCD) and the strong interacting matter possesses rich phase structure in the condition of finite temperature and density. By creating the circumstance of high temperature and high density from the Relativistic Heavy Ion  Collision (RHIC) experiments, one can investigate the QCD phase transitions which are not only important to realize the QCD phase structure, but also are critical to understand the evolution of the early universe and the internal structure of quark stars ~\cite{Shuryak:2014zxa,Rischke:2003mt,Fukushima:2010bq,Huang:2023ogw}. At low temperatures and densities, the strongly interacting matter is in the hadronic phase. The transition from hadronic phase to quark-gluon plasma (QGP) phase, namely the deconfinement transition, takes place with the increasing of temperature and chemical potential. Besides, the transition from the chiral symmetry broken phase to the chiral symmetry restoration phase occurs with the rise of temperature and chemical potential.

Since the created fireballs in RHICs last for very short time, the detection of the properties of the hot and dense medium is mainly based on the detection of the final particles, among which the hadrons play important roles. In order to make a good explanation to the experimental data from RHICs, it is essential to study the in-medium properties of hadrons which might have significant impacts on the final distribution of hadrons~\cite{Andronic:2008gu}. Furthermore, understanding the properties of hadrons under the extreme conditions is of scientific merit to reveal the phase structures of strongly interacting.

One of the most important quantities to characterize the properties of meson is the meson mass, the thermal and 
dense behaviors of which are of significance to understand the properties of hot and dense nuclear matter. 
Due to the breaking of the Lorentz symmetry at finite temperature, one can define two different kinds of meson mass in medium, namely, screening mass and pole mass.

Defined as the exponential decay of the spatial correlators, the screening mass encodes the information of spatial correlation function of meson field. Quantitatively, the screening mass is defined by the pole of spatial correlation function in the momentum space, i.e., $G^{-1}(\boldsymbol{p})|_{\boldsymbol{p}^{2}=-m_{\rm{scr}}^{2}}=0$~\cite{Ishii:2016dln,Florkowski:1997pi,Cheng:2010fe}. Physically, the inverse of screening mass, an characteristic spatial distance, can describe the screening effect that a test hadron put inside the hot medium can be effectively screened beyond this spatial distance~\cite{Cheng:2010fe}.  

The pole mass is defined by the pole of the real part of the temporal correlator $G(\omega)$ in the frequency space. Physically, the pole mass depicts the natural oscillation frequency of the particle. At $T=0$, the screening mass is equal to the pole mass because of the Lorentz invariance of the system. However, the screening mass and the pole mass are different at $T>0$ because the Lorentz invariance is broken by the existence of heat bath reference frame~\cite{Shuryak:1990ie,Pisarski:1996mt,Cao:2022csq}. 

Apart form the masses mentioned above, the thermal width, which is defined by the pole of the imaginary part of the temporal correlator $G(\omega)$ and interpreted as resonance absorption in a hot and dense nuclear matter, is also an important quantity to characterize the properties of mesons in-medium. The thermal width of meson has an important effect in RHICs. For example, the temperature dependence of the thermal width of the $\rho$ meson is of significance to measure the dilepton production in the heavy ion collision~\cite{Ayala:2012ch}. Besides, a monotonically increasing mesonic width with increasing $T$ can be related to a signal of deconfinement transition~\cite{Dominguez:2010mx,Dominguez:2013fca}.

Among the light mesons, of particular interest is the pion, known as the lightest meson as well as pseudo-Goldstone boson, which has attracted many attentions in recent years. There are several reasons to investigate the in-medium properties of pion. Firstly, it is the lightest meson so that it can reach thermal equilibrium with the medium. Furthermore, pion has a closed relationship with chiral phase transition. For example, in chiral limit, the mass of pion will vanish and the mass of scalar meson is nonzero below the chiral phase transition temperature $T_{c}$. However, above $T_{c}$, the mass of pion will become nonzero and increase with temperature and gets degenerate with scalar meson, which indicates that the chiral phase transition from chiral symmetry breaking phase to chiral symmetry restoration phase appears. What is more, at hadronic spectrum level, pion in isospin medium is also a probe for pion superfluid phase transition. When the isospin chemical potential $\mu_{I}$ grows to $m_{\pi}$ at zero temperature, the U(1) symmetry is broken spontaneously and the pion superfluid phase occurs~\cite{Migdal:1971cu}. The study of the isospin behavior of pion remains an interesting topic in hadronic physics. On the one hand, the isospin density effect can be verified directly by the lattice simulation without serious technical problems~\cite{Kogut:2002zg,Kogut:2004zg}. On the other hand, the Goldstone mode corresponding to the global isospin symmetry breaking plays a leading role on the dynamic and thermal properties of the pion superfluidity~\cite{Mu:2009ww}. 

The physics of the pion at finite temperature and density, however, is probably strong coupling and then one has to resort to the nonperturbative methods. Lattice QCD (LQCD) simulation~\cite{Ishii:2016dln,Brandt:2014qqa,Brandt:2015sxa}, as a first-principles calculation, can work very well under the finite temperature. But LQCD is complicated at finite chemical potential due to the sign problem of the fermion determinant~\cite{Fodor:2001au}. Other low energy effective models are constructed to describe the properties of pions, such as the chiral perturbation theory~($\chi $PT)~\cite{Son:2001ff,Son:2002ci}, the functional renormalization group~(FRG)~\cite{Tripolt:2013jra,Wang:2017vis}, Dyson-Schwinger Equation~(DSE)~\cite{Fischer:2018sdj,Gao:2020hwo} and the Nambu-Jona-Lasinio models~(NJL)~\cite{Ebert:1992jx,Xia:2013caa,Chao:2018ejd,Liu:2018zag,Xu:2020yag,Sheng:2021evj}. Different methods obtained the same conclusion that the pion masses increase with the increasing temperature above the chiral transition temperature $T_{c}$. However, the temperature behavior of the pion pole mass with a physical quark mass below $T_{c}$ is still controversial now. Son and Stephanov argued that $m_{\pi,{\rm{pole}}}$ decreases with the increase of temperature below $T_{c}$ in the Refs.~\cite{Son:2001ff,Son:2002ci}. This argument was supported by LQCD~\cite{Brandt:2014qqa,Brandt:2015sxa} and the NJL model with gluon condensation~\cite{Ebert:1992jx}. However, using other methods including FRG~\cite{Tripolt:2013jra}, NJL models~\cite{Xia:2013caa}, LQCD~\cite{Ishii:2016dln}, and DSEs~\cite{Gao:2020hwo} obtained the opposite result that $m_{{\rm{pole}},\pi}$ increases with the increase of temperature below $T_{c}$. Therefore, it is necessary to use other methods to study this problem. 

Developed from the anti-de Sitter/conformal field theory(AdS/CFT) correspondence~\cite{Maldacena:1997re,Gubser:1998bc,Witten:1998qj}, fortunately, the holographic methods provide an alternative robust approach to deal with the strong coupling problem of QCD~\cite{Casalderrey-Solana:2011dxg}. There are lots of useful models in the framework of bottom-up approach, such as the hard wall model~\cite{Erlich:2005qh}, the soft wall model~\cite{Karch:2006pv}, the light-front holographic QCD~\cite{Brodsky:2014yha} and the Einstein-Maxwell-Dilaton model~\cite{Gubser:2008ny,Gubser:2008yx,DeWolfe:2010he,Gursoy:2007cb,Gursoy:2007er}. Among these models, the soft-wall AdS/QCD model and its extended models  give well description of the chiral phase  transition~\cite{Cherman:2008eh,Gherghetta:2009ac,Li:2012ay,Colangelo:2011sr,Li:2016smq,Chelabi:2015gpc,Chelabi:2015cwn,Fang:2016nfj,Chen:2018msc,Ballon-Bayona:2020qpq,Ballon-Bayona:2021ibm,Chen:2021wzj}. These models also can well describe the glueball and hadron spectra~\cite{Colangelo:2008us,Sui:2009xe,Kelley:2010mu,Li:2013oda,FolcoCapossoli:2015jnm,Dudal:2015wfn,FolcoCapossoli:2019imm,Cao:2020ryx,Cao:2021tcr,Chen:2019rez,Zhao:2021ogc,Mamani:2022qnf,Guo:2023zjx}. Consequently, we would like to investigate the pion spectra in the framework of soft-wall AdS/QCD model. 

 There are many efforts have been made to investigate the isospin behaviors of pion in the hard-wall model~\cite{Lee:2013oya,Nishihara:2014nva,Nishihara:2014nsa,Mamedov:2015sha,Nasibova:2023fyh} and the soft-wall model~\cite{Lv:2018wfq,Cao:2020ske}. However, most of these literatures considered temperature effect and isospin density effect separately. It is meaningful to consider both of them at the same time and study the mutual effects for the pion spectra. Based on the soft-wall AdS/QCD model, there are some investigations on the pion pole mass and screening mass and their thermal properties at finite temperature and isospin chemical potential~\cite{Cao:2020ryx,Cao:2021tcr}  through the spectral function method~\footnote{The spectral function method has been widely used in the studies in holographic QCD models~\cite{Teaney:2006nc,Kovtun:2006pf,Colangelo:2009ra,Colangelo:2012jy}.  Furthermore, the spectral function can also extracted from the lattice data from the spatial correlator~\cite{Lowdon:2022xcl}}. As the temperature rises, however, it is difficult to determine the location of the resonance peak of the spectrum functions because the resonance peak gets inconspicuous. Therefore, in this paper, we resort to another method by calculating the quasi-normal mode(QNM), of which the real part denotes the pole mass and the imaginary part denotes the thermal width~\cite{Kovtun:2005ev,Miranda:2009uw}. With the QNM method, we extend the studies~\cite{Cao:2021tcr} to finite $\mu_{I}$ and investigate the isospin behavior of screening mass as well as its relation with the pion superfluid phase transition.

The paper is organized as follows. In Sec.~\ref{sec2}, we will give a brief review of the soft-wall AdS/QCD model. In Sec.~\ref{sec3} we extract the screening mass of pion  quasiparticles by calculate the poles of the spatial  correlation functions at finite temperature and isospin chemical potential. We will also study the temporal correlation functions and extract the pole masses and thermal widths from QNMs. In Sec.~\ref{sec4} we give our conclusion and summary.

\section{\label{sec2}Soft-wall AdS/QCD models with finite isospin chemical potential}
In the bottom-up approach, the soft-wall AdS/QCD model~\cite{Karch:2006pv} can describe both spontaneously chiral symmetry breaking and linear confinement in the vacuum qualitatively. Here, we review the soft-wall AdS/QCD model briefly.

The action of $N_{f}=2$ soft-wall AdS/QCD model constructed with the $\rm{SU}_{L}(2)\times\rm{SU}_{R}(2)$ gauge symmetry under the dual 5D geometry~\cite{Karch:2006pv} takes the following form
\begin{eqnarray}
S=&&\int d^{4}x\int dz\sqrt{g}e^{-\Phi} {\rm{Tr}} 
\left\{
  \left|D_{M}X\right|^{2}-V(\left|X\right|)\right.\nonumber \\&& \left.-\frac{1}{4g_{5}^{2}}(F_{L}^{2}+F_{R}^{2})
  \right\}, 
  \label{eq:one}
\end{eqnarray}
where $g$ is the determinant of the metric $g_{MN}$. $\Phi(z)=\mu_{g}^2z^2$ is the quadratic dilaton field which depends on the fifth dimension $z$~\cite{Karch:2006pv}. The gauge coupling constant is $g_{5}=2\pi$, when the number of colors is $N_{c}=3$ by comparing the vector current two-point function in large-momentum expansion to the large-$N_{c}$ QCD perturbative result~\cite{Erlich:2005qh}. We will take $N_{c}=3$ in the following calculation. $X$ is the matrix-valued bulk scalar field, and the covariant derivative $D_{M}X$ with $M=(x,z)$ is defined as
\begin{equation}
    D_{M}X=\partial_{M}X-iL_{M}X+iXR_{M},
\end{equation}
where $L_{M}$ and $R_{M}$ are the chiral gauge fields,
\begin{equation}
    L_{M}=L_{M}^{a}t^{a},\qquad R_{M}=R_{M}^{a}t^{a}.
\end{equation}
$t^{a}=\sigma^{a}/2(a=1,2,3)$ are the generators of $\rm{SU}(2)$. The potential term takes
\begin{equation}
    V(\left|X\right|)=m_{5}^{2}\left|X\right|^{2}+\lambda\left|X\right|^{4},
\end{equation}
with $m_{5}^{2}(z)$ the modified 5D mass ~\cite{Fang:2016nfj} and  $\lambda$ a free parameter. $F_{MN}^{L/R}$ are the field strength tensors of the corresponding chiral gauge fields, which are defined by
\begin{subequations}
\begin{equation}
F_{MN}^{L}=\partial_{M}L_{N}-\partial_{N}L_{M}-i[L_{M},L_{N}],
\end{equation}
\begin{equation}
 F_{MN}^{R}=\partial_{M}R_{N}-\partial_{N}R_{M}-i[R_{M},R_{N}].
\end{equation}
\end{subequations}

For convenience, we can redefine the chiral gauge fields as the vector gauge field and the axial-vector gauge field
\begin{subequations}
\label{eq:6}
\begin{equation}
 V_{M}=\frac{L_{M}+R_{M}}{2},\label{subeq:1}
\end{equation}
\begin{equation}
 A_{M}=\frac{L_{M}-R_{M}}{2},\label{subeq:2}
\end{equation}
\end{subequations}
where the vector field $V_{M}$ and the axial-vector field $A_{M}$ dual to the vector current $J_{\mu}^V$ and axial-vector current $J_{\mu}^A$, respectively. For example, the isospin current $\overline{q}\gamma_{\mu}t^{3}q$ is dual to $V_{\mu}^{3}$. After the transformation in Eqs.~\eqref{eq:6}, we obtain the gauge fields strengths
\begin{subequations}
\begin{equation}
F_{MN}^{V}=\partial_{M}V_{N}-\partial_{N}V_{M}-i[V_{M},V_{N}]-i[A_{M},A_{N}],
\end{equation}
\begin{equation}
  F_{MN}^{A}=\partial_{M}A_{N}-\partial_{N}A_{M}-i[V_{M},A_{N}]-i[A_{M},V_{N}],
\end{equation}
\end{subequations}
and the covariant derivative 
\begin{equation}
    D_{M}X=\partial_{M}X-i[V_{M},X]-i\{A_{M},X\}.
\end{equation}

We consider the temperature as well as the isospin chemical potential effect and take the following metric ansatz
\begin{equation}
    ds^{2}=e^{2A(z)}\left[f(z)dt^{2}-d\boldsymbol{x}^{2}-\frac{1}{f(z)}dz^{2}\right],
    \label{eq:9}
\end{equation}
If there is a horizon $z=z_{h}$ where $f(z)=0$, one can define the temperature by the following formula
\begin{equation}
    T=\frac{\lvert f'(z_{h})\rvert}{4\pi}.\label{eq:10}
\end{equation}
According to the holographic dictionary, the conserved current is dual to the gauge field in Eq.~(\ref{subeq:1}). In general, $A(z)$ and $f(z)$ should be solved from a certain kind of gravity system which is coupled with the soft-wall AdS/QCD model action. For simplicity, we calculate in the sense of probe limit. We consider the Anti-de Sitter-Reissner-Nordstrom (AdS-RN) metric solution with finite isospin chemical potential
\begin{equation}
    A(z)=-\ln(z),\label{eq:11}
\end{equation}
\begin{equation}
    f(z)=1-(1+\mu_{I}^{2}z_{h}^{2})\frac{z^{4}}{z_{h}^{4}}+\mu_{I}^{2}\frac{z^{6}}{z_{h}^{4}}\label{eq:12}
\end{equation}
with  $\mu_{I}$ the isospin chemical potential. $V_{0}^{3}$ satisfies the following form
\begin{equation}
\label{eq13}
    V_{0}^{3}(z)=\mu_{I}\Big(1-\frac{z^{2}}{z_{h}^{2}}\Big).
\end{equation} 
 For convenience, we denote $V_{0}^{3}(z)$ by $\nu(z)$. From Eqs.~\eqref{eq:10} and~\eqref{eq:12}, we can obtain the temperature 
\begin{equation}
    T=\frac{2-\mu_{I}^{2}z_{h}^{2}}{2\pi z_{h}}.
\end{equation}

In this work, we only consider two lighteset flavors of quarks, namely up (u) quark and down (d) quark, with equal masses $m_{q}=m_{u}=m_{d}$. Then we get the matrix-valued scalar field
\begin{equation}
    X=\frac{\chi}{2}\rm{I}.
    \label{eq:15}
\end{equation}
Here, $\rm{I}$ is the two dimension identity matrix. Inserting Eqs.~\eqref{eq:9} and~\eqref{eq:15} into the 5D action Eq.~\eqref{eq:one}, we can obtain the equation of motion (EOM) of $\chi$ as follow
\begin{eqnarray}
\label{eq16}
    \chi''&&+\left(3A'-\Phi'+\frac{f'}{f}\right)\chi'-\frac{e^{2A}}{f}\Big(m_{5}^{2}\chi+\frac{\lambda\chi^{3}}{2}\Big)=0.~~~~
\end{eqnarray}

By solving the EOM of $\chi$, one can obtain the temperature and isospin chemical potential dependent behavior of chiral condensate. However, it is a second-order nonlinear ordinary differential equation and it is hard to obtain the analytical solution. Therefore, we must resort to the numerical solutions. 

\begin{figure*}[htb]
    \centering
    \begin{overpic}[width=0.42\textwidth]{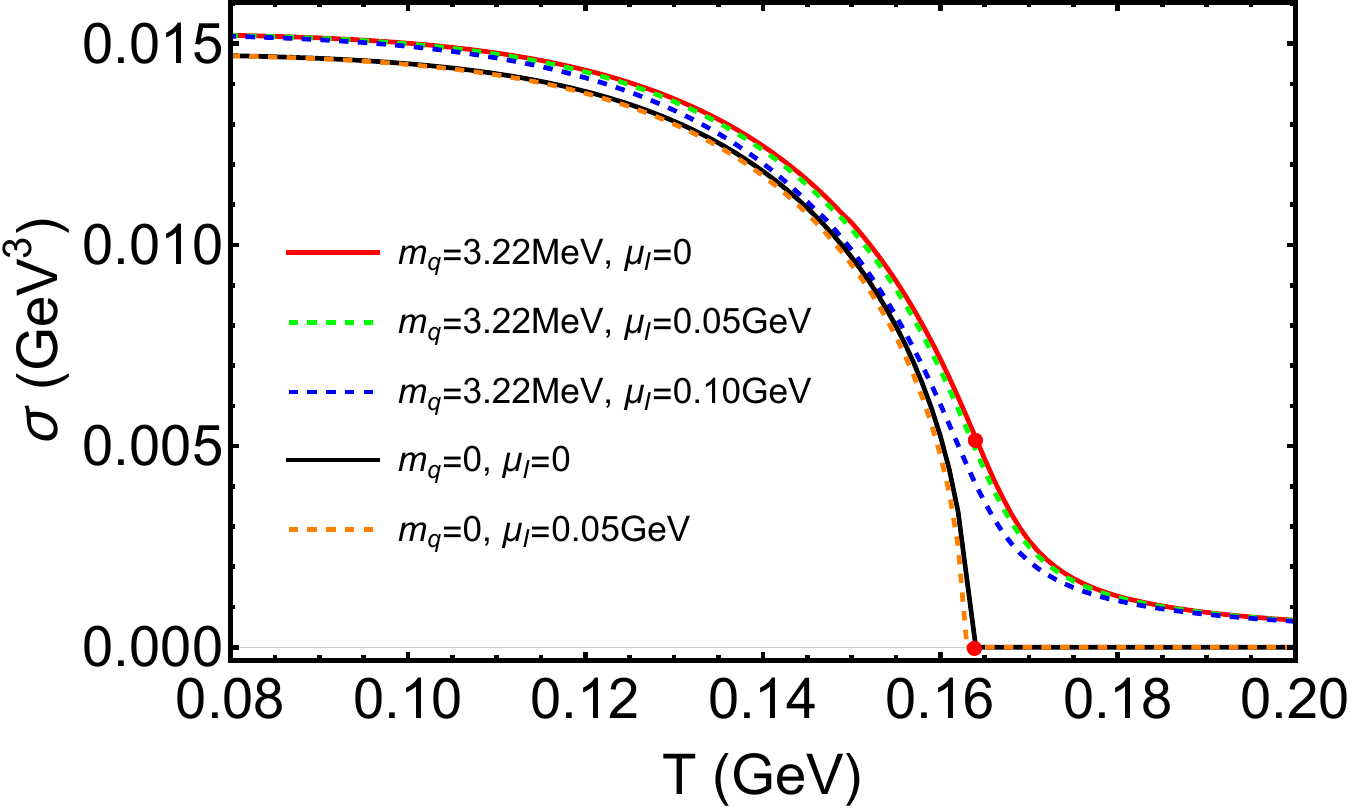}
        \put(85,50){\bf{(a)}}
    \end{overpic}
    \begin{overpic}[width=0.42\textwidth]{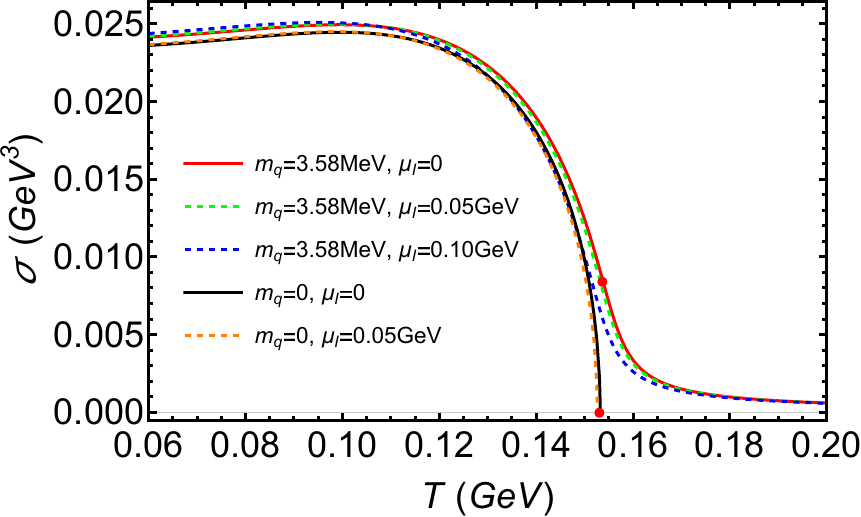}
        \put(85,50){\bf{(b)}}
    \end{overpic}
    \caption{\textbf{(a)} The chiral condensate $\sigma$ of model I as a function of temperature $T$ with different isospin chemical potential $\mu_{I}$ in chiral limit ($m_{q}=0$) and with the physical quark mass ($m_{q}=3.22$ MeV), respectively. In chiral limit, $\sigma$ vanishes at the critical temperature $T_{c}$ (at $\mu_{I}=0$, $T_{c}\approx 0.1633$ GeV). With finite physical quark mass, however, the second-order phase transition turns to a crossover with a pseudo-critical temperature $T_{cp}\approx 0.1639$ GeV at $\mu_{I}=0$ (As shown by the red dots). \textbf{(b)} The chiral condensate $\sigma$ of model II as a function of temperature $T$ with different isospin chemical potential $\mu_{I}$ in chiral limit ($m_{q}=0$) and with the physical quark mass ($m_{q}=3.58$ MeV). The red points in (b) stand for the critical temperature $T_{c}\approx 0.1532$ GeV and the pseudo-critical temperature $T_{cp}\approx 0.1537$ GeV at $\mu_{I}=0$, respectively.}
    \label{fig:1}
\end{figure*}

To obtain general features of the soft-wall AdS/QCD models, herein, we consider two kinds of soft-wall AdS/QCD models with different modified 5D masses $m_{5}^2(z)$ which are introduced to obtain a good description of both chiral symmetry spontaneously breaking and meson spectrum. The modified forms of $m_5^2(z)$ are shown in Table~\ref{tab:table1}. Model I is introduced in Ref.~\cite{Fang:2016nfj}. In model II, we consider the modification of $m_5^2(z)$ as the coupling to the dilaton $\Phi(z)$.
\begin{table}[htbp]
\centering
\caption{\label{tab:table1}%
Two kinds of soft-wall AdS/QCD models with different 5D masses $m_{5}^2(z)$.
}
\begin{ruledtabular}
\begin{tabular}{ccc}
Model&I&II\\ 
\hline
\\
$m_{5}^2(z)$ & $-3-\mu_{c}^2z^2$ & $-3[1+\gamma \tanh(\kappa\Phi)]$ 
\\
\end{tabular}
\end{ruledtabular}
\end{table}

For model I, one can obtain the asymptotic behaviors at UV boundary ($z=0$) and horizon ($z=z_{h}$)
\begin{subequations}
\label{eq:whole1}
\begin{eqnarray}
\chi(z\rightarrow0)&&=m_{q}\zeta z+\frac{\sigma}{\zeta}z^{3}+\frac{m_{q}\zeta}{4}\Big(-2\mu_{c}^{2}\nonumber\\
&&+4\mu_{g}^{2}+m_{q}^{2}\zeta^{2}\lambda\Big)z^{3}\ln(z)+O(z^{4})
\end{eqnarray}
\begin{eqnarray}
  \chi(z\rightarrow z_{h})&&=c_{0}+\frac{c_{0}(2\mu_{c}^{2}z_{h}^{2}-c_{0}^{2}\lambda+6)}{8z_{h}-4z_{h}^{3}\mu_{I}^{2}}(z-z_{h})\nonumber\\
  &&+O[(z-z_{h})^{2}]
\end{eqnarray}
\end{subequations}
where the two independent integral constants $m_{q}$ and $\sigma$ in the UV are dual to the quark mass and chiral condensate $\sigma\equiv \left\langle \overline{q}q\right\rangle$, respectively, according to the holographic dictionary. Here, $\zeta$ is a normalization constant which equals $\sqrt{N_{c}}/2\pi$, by matching the correlations of $\bar{q}q$ operator to 4D results~\cite{Cherman:2008eh}. Furthermore, $c_{0}$ is the integral constant generating a regular solution at the horizon. 
For model II, we can also obtain the asymptotic series at at UV boundary ($z=0$) and horizon ($z=z_{h}$)
\begin{subequations}
\label{eq:whole2}
\begin{eqnarray}
 \chi(z\rightarrow0)&&=m_{q}\zeta z+\frac{\sigma}{\zeta}z^{3}+\frac{1}{4}\Big[m_{q}(4-6\gamma\kappa)\mu_{g}^2\zeta\nonumber\\
 &&+m_{q}^3\lambda\zeta^3\Big]z^3\ln(z)+O(z^{4})
\end{eqnarray}
\begin{eqnarray}
  \chi(z\rightarrow z_{h})&&=c_{0}+\frac{c_{0}\Big[6-c_{0}^2\lambda+6\gamma \tanh(z_{h}^2\kappa \mu_{g}^2)\Big]}{8z_{h}-4z_{h}^3\mu_{I}^2}\nonumber\\
  &&\times(z-z_{h})+O[(z-z_{h})^{2}]
\end{eqnarray}
\end{subequations}
We adopt the parameters for model I according to Ref.~\cite{Fang:2016nfj}, which are shown in the Table~\ref{tab:table2}. As for model II, we adopt the parameters shown in the Table~\ref{tab:table3}, which are determined by fitting the meson spectrum. 
\begin{table}[htbp]
\caption{\label{tab:table2}%
Parameters in model I.
}
\begin{ruledtabular}
\begin{tabular}{ccccc}
\textrm{Parameters}&
\textrm{$m_{q}$(GeV)}&
\multicolumn{1}{c}{\textrm{$\mu_{g}$(GeV)}}&
\textrm{$\mu_{c}$(GeV)}&
\textrm{$\lambda$}\\
\colrule
Value  & $3.22\times10^{-3}$ & 0.44 &1.45 & 80 \\
\end{tabular}
\end{ruledtabular}
\end{table}

\begin{table}[htbp]
\caption{\label{tab:table3}%
Parameters in model II.
}
\begin{ruledtabular}
\begin{tabular}{cccccc}
\textrm{Parameters}&
\textrm{$m_{q}$(GeV)}&
\multicolumn{1}{c}{\textrm{$\mu_{g}$(GeV)}}&
\textrm{$\gamma$}&
\textrm{$\lambda$}&
\textrm{$\kappa$}\\
\colrule
Value  & $3.58\times10^{-3}$ & 0.43 &3.7 & 14.7 &1 \\
\end{tabular}
\end{ruledtabular}
\end{table}

With the boundaries conditions Eqs.~\eqref{eq:whole1} and \eqref{eq:whole2}, we can solve the Eq.~\eqref{eq16} by ``shooting method''~\cite{boyd2001chebyshev} and extract the chiral condensate $\sigma$ as a function of temperature $T$ or isospin chemical potential $\mu_{I}$. The  relevant numerical results are presented in Fig.~\ref{fig:1}. 

In model I, the result show that the chiral condensate $\sigma$ is decreasing monotonously with the increasing of temperature in chiral limit and with physical quark mass. In chiral limit, $\sigma$ vanishes at a critical temperature $T_{c}$ (at $\mu_{I}=0$, $T_{c}\approx0.163$ GeV). With finite physical quark mass and zero isospin chemical potential, $m_q=3.22$ MeV, however, the second-order phase transition turns to a crossover with a pseudo-critical temperature $T_{cp}\approx0.164$ GeV. \footnote{ The pseudo-critical temperature $T_{cp}$ is defined by $d^2\sigma(T)/dT^2|_{T=T_{cp}}=0$.} In model II, the critical temperature $T_{c}\approx 0.1532$ GeV in the chiral limit. The pseudo-critical temperature $T_{cp}\approx 0.1537$ GeV at $\mu_{I}=0$ and $m_q=3.58$ MeV. We found that with the increasing of $\mu_{I}$, the curve of $\sigma$ shifts towards the sigma axis, which suggests the fact that isospin chemical potential tends to destroys the chiral symmetry.

\section{\label{sec3}correlation functions and mass of pions at finite temperature and isospin density}
In the last section, we have briefly reviewed the soft-wall AdS/QCD model and obtained the temperature dependent behavior of chiral condensate at different $\mu_{I}$. In this section, we will calculate screening masses and pole masses,  as well as thermal widths of pions at finite isospin density and temperature, from which one can obtain the information of pion superfluid phase transition at finite temperature.  

The screening mass $m_{\rm{scr}}$ is defined as the exponential decay of spatial correlator, i.e. the inverse of the correlation length $\xi\sim 1/m_{\rm{scr}}$. Transforming into the momentum space, it corresponds to  the pole of the Retarded correlator,
\begin{equation}
\label{eq19}
    G(\boldsymbol{p})\sim\frac{1}{\boldsymbol{p}^{2}+m_{\rm{scr}}^{2}}.
\end{equation}
As for the pole mass $m_{\rm{pole}}$ and the thermal width $\Gamma$, they are the real and imaginary part of frequency ($\omega_{0}=m_{\rm{pole}}-i\Gamma/2$) of the corresponding QNM, which is the pole of the temporal Retarded correlator in the frequency space,
\begin{equation}
    G(\omega)\sim\frac{1}{\omega-(m_{\rm{pole}}-i\Gamma/2)}.
\end{equation}

Holographic approach, connecting the 4D operator $\hat{O}(x)$ and 5D field $\phi(x,z)$ through the equivalence of the partition functions, provides a powerful tool to calculate the strong coupling correlation function, namely
\begin{equation}
    \left\langle e^{i \int d^{4}x\phi_{0}(x)\hat{O}(x)}\right\rangle=e^{iS_{5D}[\phi]}|_{\phi(x,z=0)=\phi_{0}(x)},
\end{equation}
where $\phi$ is the classical solution of the 5D action $S_{5D}$ and its boundary value $\phi(x,z=0)$ equals the 4D external source $\phi_{0}(x)$~\cite{Maldacena:1997re,Gubser:1998bc,Witten:1998qj}. By taking second derivative of the action $S_{5D}$ with respect to the source $\phi_{0}$, one can obtain the correlator $\small\left\langle \hat{O}(x)\hat{O}(0)\small\right\rangle$~\cite{Son:2002sd}.

\subsection{Pseudo-scalar channel}
In this part, we will derive the spatial correlation functions as well as the temporal correlation functions for the pseudo-scalar meson. In 4D quantum field theory, the particles are recognized as the excitation modes of the vacuum, while they are the perturbations on the background fields in the dual 5D gravity theory. For the pions, we have
\begin{equation}
\label{eq22}
    X=\frac{\rm{I}}{2}\chi  e^{2i\pi^{a}t^{a}},
\end{equation}
where $\rm{I}$ is a two dimensional identity matrix and $\pi^{a}$ (a=1,2,3) is the pion perturbation. Here, we have neglect other channel perturbations which do not affect our discussion. Substituting Eq.~\eqref{eq22} into the Eq.~\eqref{eq:one} and keeping to the quadratic terms, together with the gauge condition $A_z=0$, one can obtain the action of pion part as
\begin{eqnarray}\label{pionaction}
    S_{PS}&&=\int d^4x\int_{0}^{z_{h}}dz\sqrt{g}e^{-\Phi}
       \Bigg\{
       \frac{1}{2}(M_{A}^2)_{ab}\Big[g^{zz}\partial_{z}\pi^{a}\partial_{z}\pi^{b}\nonumber\\
&&+g^{\mu\nu}\partial_{\mu}\pi^{a}\partial_{\nu}\pi^{b}-2g^{\mu\nu}\partial_{\mu}\pi^{a}A_{\nu}^{b}+g^{\mu\nu}A_{\mu}^{a}A_{\nu}^{b}\Big]\nonumber\\
       &&+g^{tt}
      \Big[\frac{1}{2}\nu(z)^{2}(M_{D}^2)_{ab}\pi^a\pi^b+\nu(z)(M_{I}^2)_{ab}
      (\pi^{b}\partial_{t}\pi^{a}\nonumber\\
      &&+\pi^{a}A_{t}^{b}
      )
      \Big]-\frac{1}{2g_{5}^2}g^{zz}g^{\mu\nu}\partial_{z}A_{\mu}^{a}\partial_{z}A_{\nu}^{a}-\frac{1}{2g_{5}^2}g^{tt}g^{ii}\nonumber\\
      &&\times(\partial_{t}A_{i}^a-\partial_{i}A_{t}^a)^2
     \Bigg\},
\end{eqnarray}
where $(M_{A}^2)_{ab}$, $(M_{I}^2)_{ab}$ and $(M_{D}^2)_{ab}$ are $3\times3$ matrices defined as follows, 
\begin{subequations}
\begin{eqnarray}
    (M_{A}^2)_{ab}=\begin{pmatrix}
    \chi^2 & 0 & 0 \\
    0 & \chi^2 & 0 \\
    0 & 0 & \chi^2 \\
    \end{pmatrix},
\end{eqnarray}
\begin{eqnarray}
    (M_{I}^2)_{ab}=\begin{pmatrix}
    0 & -\chi^2 & 0 \\
    \chi^2 & 0 & 0 \\
    0 & 0 & 0 \\
    \end{pmatrix},
\end{eqnarray}
 \begin{eqnarray}
    (M_{D}^2)_{ab}=\begin{pmatrix}
    \chi^2 & 0 & 0 \\
    0 & \chi^2 & 0 \\
    0 & 0 & 0 \\
    \end{pmatrix}, 
 \end{eqnarray}  
\end{subequations}
with a,b generator indexes of $\rm{SU}(2)$.
Here, $\pi^{a}$ and $A_{\mu}^{a}$ are functions of the coordinates $x=(t,-\boldsymbol{x})$ and $z$. By taking the Fourier transformation, 
\begin{subequations}
    \begin{eqnarray}
        \pi^{a}(x,z)=\frac{1}{(2\pi)^{4}}\int d^4k e^{i kx}\pi^{a}(k,z),
    \end{eqnarray}
\begin{eqnarray}
     A_{\mu}^{a}(x,z)=\frac{1}{(2\pi)^{4}}\int d^4ke^{i kx} A_{\mu}^{a}(k,z),
\end{eqnarray}
\end{subequations}
one can solve the equation of motions in the momentum space $k=(\omega,- \boldsymbol{p})$ .
Without losing generality and for simplicity, we assign $\boldsymbol{p}$ along the $x_{1}$-direction, i.e., $\boldsymbol{p}=(p,0,0)$. Since the  breaking of the isospin symmetry at finite isospin chemical potential, the neutral pion $\pi^3$ and the charged pions $\pi^{1,2}$ would no longer be degenerate. Thus, we have to take the isospin index ($a=1,2,3$) into account. For convinence, we define $\pi^3=\pi^0$, and take a rotation in the isospin space

\begin{subequations}
    \begin{eqnarray}
        \begin{pmatrix}
        \pi^1\\
        \pi^2\\
    \end{pmatrix}=\begin{pmatrix}
    \frac{1}{\sqrt{2}} & \frac{1}{\sqrt{2}}  \\
    \frac{i}{\sqrt{2}} &  -\frac{i}{\sqrt{2}} \\
    \end{pmatrix}\begin{pmatrix}
        \pi^+\\
        \pi^-\\
    \end{pmatrix},
    \end{eqnarray}
    \begin{eqnarray}
         \begin{pmatrix}
        A_{t}^1\\
        A_{t}^2\\
    \end{pmatrix}=\begin{pmatrix}
    \frac{1}{\sqrt{2}} & \frac{1}{\sqrt{2}}  \\
    \frac{i}{\sqrt{2}} &  -\frac{i}{\sqrt{2}} \\
    \end{pmatrix}\begin{pmatrix}
        A_{t}^+\\
        A_{t}^-\\
    \end{pmatrix}.
    \end{eqnarray}
\end{subequations}
From the action in Eq.~\eqref{pionaction}, the EOMs of $\pi^0, A_{t}^0$,  and $A_{i}^0$, are derived as
\begin{subequations}
\label{eq29}
    \begin{eqnarray}
       &&\pi^{0}{''}+\left(3A'-\Phi'+\frac{f'}{f}+2\frac{\chi'}{\chi}\right)\pi^{0}{'}\nonumber\\
    &&+\left(\frac{\omega^2}{f^2}-\frac{p^2}{f}\right)\pi^0-\frac{i\omega}{f^2}A_{t}^0-\frac{ip}{f}A_{i}^0=0, \label{eq29a}
    \end{eqnarray}
\begin{eqnarray}
     &&A_{t}^{0}{''}+(A'-\Phi')A_{t}^{0}{'}-\frac{g_{5}^2e^{2A}\chi^2 i\omega}{f}\pi^0\nonumber\\
     &&-\frac{p^2+g_{5}^2e^{2A}\chi^2 }{f}A_{t}^0-\frac{\omega p}{f}A_{i}^0=0,
     \label{eq29b}
\end{eqnarray}
\begin{eqnarray}
    &&A_{i}^{0}{''}+\Big(A'-\Phi'+\frac{f'}{f}\Big)A_{i}^{0}{'}+\frac{g_{5}^2e^{2A}\chi^2 ip}{f}\pi^0\nonumber\\
    &&+\frac{\omega^2-fg_{5}^2e^{2A}\chi^2 }{f^2}A_{i}^0+\frac{\omega p}{f^2}A_{t}^0=0,\label{eq29c}
\end{eqnarray}
\end{subequations}
and the EOMs of $\pi^{\pm}, A_{t}^{\pm}$, and $A_{i}^{\pm}$, are derived as
\begin{subequations}
\label{eq28}
\begin{eqnarray}
  &&\pi^{\pm}{''}+\Big(3A'-\Phi'+\frac{f'}{f}+2\frac{\chi'}{\chi}\Big)\pi^{\pm}{'}\nonumber\\
    &&+\frac{(\omega \pm\nu)^2-fp^2}{f^2}\pi^{\pm}-\frac{i(\omega \pm \nu)}{f^2}A_{t}^{\pm}\nonumber\\
    &&-\frac{ip}{f}A_{i}^{\pm}=0, \label{eq29d}
    \end{eqnarray}
\begin{eqnarray}
    &&A_{t}^{\pm}{''}+(A'-\Phi')A_{t}^{\pm}{'}-\frac{g_{5}^2e^{2A}\chi^2 i(\omega\pm\nu)}{f}\pi^{\pm}\nonumber\\
    &&-\frac{p^2+g_{5}^2e^{2A}\chi^2 }{f}A_{t}^{\pm}-\frac{\omega p}{f}A_{i}^{\pm}=0,\label{eq29e}
\end{eqnarray}
\begin{eqnarray}
     &&A_{i}^{\pm}{''}+\Big(A'-\Phi'+\frac{f'}{f}\Big)A_{i}^{\pm}{'}+\frac{g_{5}^2e^{2A}\chi^2 ip}{f}\pi^{\pm}\nonumber\\
    &&+\frac{\omega^2-fg_{5}^2e^{2A}\chi^2 }{f^2}A_{i}^{\pm}+\frac{\omega p}{f^2}A_{t}^{\pm}=0,\label{eq29f}
\end{eqnarray}
\end{subequations}
where the prime represents the derivative with respect to $z$.

Note that the EOMs of pion fields are coupled linear second-order differential equations with two singularities. The analytical solutions are almost impossible to get. However, one can solve them numerically. We can get the asymptotic expansions of $\pi^0$ , $A_{t}^0$ and $A_{i}^0$ at the UV boundary,
\begin{subequations}
\label{eq30}
    \begin{eqnarray}
         \pi^0(z\rightarrow0)&=&\pi_{0}+\frac{1}{2}\Big[
    \pi_{0}(p^2-\omega^2)+i (\varphi_{0}p+a_{0}\omega)
    \Big]\nonumber\\
    &&\times z^2\ln(z)+\pi_{2}z^2+\mathcal{O}(z^3),
    \end{eqnarray}
    \begin{eqnarray}
        A_{t}^0(z\rightarrow0)&=&a_{0}+\frac{1}{2}\Big[
    a_{0}(p^2+g_{5}^2m_{q}^2\zeta^2)+\varphi_{0}p\omega\nonumber\\
    &&+ig_{5}^2m_{q}^2\zeta^2\pi_{0}\omega
    \Big]z^2\ln(z)+a_{2}z^2+\mathcal{O}(z^3),\nonumber\\
    \end{eqnarray}
    \begin{eqnarray}
         A_{i}^0(z\rightarrow0)&=&\varphi_{0}+\frac{1}{2}\Big[
    p(-i g_{5}^2m_{q}^2\zeta^2\pi_{0}-a_{0}\omega)\nonumber\\
    &&+\varphi_{0}(g_{5}^2m_{q}^2\zeta^2-\omega^2)
    \Big]z^2\ln(z)+\varphi_{2}z^2\nonumber\\
    & &+\mathcal{O}(z^3),
    \end{eqnarray}
\end{subequations}
where $\pi_{0}$, $\pi_{2}$, $a_{0}$, $a_{2}$, $\varphi_{0}$, $\varphi_{2}$ are integration constants. According to the holographic dictionary, $\pi_{0}$, $a_{0}$, $\varphi_{0}$ are decoded as the external source $J_{\pi}$, $J_{A_{t}}$ and $J_{A_{i}}$. At the horizon, we can also get the asymptotic expansions as~\footnote{Here, we take the in-coming wave solution and neglect the out-going one.}
\begin{subequations}
\label{eq:30}
    \begin{eqnarray}
        \pi^0(z\rightarrow z_{h})&=& (z_{h}-z)^{\frac{i \omega z_{h}}{2\mu_{I}^2z_{h}^2-4}}\Big\{\pi_{h0}+\pi_{h1}(z-z_{h})\nonumber\\
        &&+\mathcal{O}[(z-z_{h})^2]\Big\}\nonumber\\
    &&+b_{h0}+\frac{i b_{h1}z_{h}^2\omega(z-z_{h})}{16+4z_{h}^4\mu_{I}^4+z_{h}^2(-16\mu_{I}^2+\omega^2)}
   \nonumber\\
    && +\mathcal{O}[(z-z_{h})^2], \label{eq31a}
    \end{eqnarray}
    \begin{eqnarray}
        A_{t}^0(z\rightarrow z_{h})&=&(z_{h}-z)^{\frac{i \omega z_{h}}{2\mu_{I}^2z_{h}^2-4}}\Big\{a_{h1}(z-z_{h})\nonumber\\
    &&+\mathcal{O}[(z-z_{h})^2]\Big\}-i b_{h0}\omega+b_{h1}(z-z_{h})
    \nonumber\\
    &&+\mathcal{O}[(z-z_{h})^2],\label{eq31b}
    \end{eqnarray}
    \begin{eqnarray}
        A_{i}^0(z\rightarrow z_{h})&=&(z_{h}-z)^{\frac{i \omega z_{h}}{2\mu_{I}^2z_{h}^2-4}}\nonumber\\
    &&\times\Big\{-i\Big[\frac{2 \pi_ {h0} c_ {0}^2 g_ {5}^2(-2 + z_ {h}^2\mu_ {I}^2)}{2pz_{h}^2(-2 + z_ {h}^2\mu_ {I}^2)}\nonumber\\
    &&+\frac{a_ {h1} z_ {h}^2 (4 - 2 z_ {h}^2\mu_ {I}^2 - iz_ {h}\omega)}{2pz_{h}^2(-2 + z_ {h}^2\mu_ {I}^2)}\Big]\nonumber\\
    &&+\varphi_{h1}(z-z_{h})+\mathcal{O}[(z-z_{h})^2]
    \Big\}+i b_{h0}p\nonumber\\
    &&-\frac{b_{h1}pz_{h}^2\omega(z-z_{h})}{16+4z_{h}^4\mu_{I}^4+z_{h}^2(-16\mu_{I}^2+\omega^2)}\nonumber\\
    &&+\mathcal{O}[(z-z_{h})^2]\label{eq31c} 
    \end{eqnarray}   
\end{subequations}
with coefficients of first order in Eq.~\eqref{eq31a} and Eq.~\eqref{eq31c},
\begin{subequations}
\begin{eqnarray}
\pi_{h1}&=&\Big\{
-2a_{h1}z_{h}^2(-2+z_{h}^2\mu_{I}^2)^2+\pi_{h0}c_{0}^2(-2+z_{h}^2\mu_{I}^2)\nonumber\\
    &&\times\Big[2g_{5}^2(-2+z_{h}^2\mu_{I}^2)-iz_{h}\lambda\omega\Big]\nonumber\\&&+\pi_{h0}z_{h}^2\Big\{2p^2(-2+z_{h}^2\mu_{I}^2)^2+i\omega\Big[4z_{h}^5\mu_{g}^2\mu_{I}^4\nonumber\\&&+z_{h}^3\mu_{I}^2(2\mu_{c}^2-16\mu_{g}^2-3\mu_{I}^2)+z_{h}(-4\mu_{c}^2\nonumber\\&&+16\mu_{g}^2+6\mu_{I}^2)+6i\omega-9iz_{h}^2\mu_{I}\omega
    \Big]
    \Big\}
\Big\}\Big/\Big[4z_{h}(-2\nonumber\\&&+z_{h}^2\mu_{I}^2)^2(-2+z_{h}^2\mu_{I}^2+iz_{h}\omega)
\Big],
\end{eqnarray} 
\begin{eqnarray}
    \varphi_{h1}&=&i\Big\{
    -4\pi_{h0}c_{0}^2g_{5}^2p^2z_{h}^2(-2+z_{h}^2\mu_{I}^2)^3\nonumber\\&&+2ia_{h1}p^2z_{h}^5(-2+z_{h}^2\mu_{I}^2)^2\omega+\Big[-2\pi_{h0}c_{0}^2g_{5}^2(-2\nonumber\\&&+z_{h}^2\mu_{I}^2)+a_{h1}z_{h}^2(-4+2z_{h}^2\mu_{I}^2+iz_{h}\omega)
    \Big]\nonumber\\
    &&\times\Big[2c_{0}^2g_{5}^2(-2+z_{h}^2\mu_{I}^2)^2+iz_{h}\omega\Big(-4+4z_{h}^6\mu_{g}^2\mu_{I}^4\nonumber\\&&+16z_{h}^2(\mu_{g}^2+\mu_{I}^2)-z_{h}^4(16\mu_{g}^2\mu_{I}^2+7\mu_{I}^4)\nonumber\\&&+6iz_{h}\omega-9iz_{h}^3\mu_{I}^2\omega
    \Big)
    \Big]
    \Big\}\Big/\Big\{
8z_{h}^3(-2+z_{h}^3\mu_{I}^2)^4\nonumber\\&&\times\Big(p+\frac{ipz_{h}\omega}{-2+z_{h}^2\mu_{I}^2}
    \Big)
    \Big\},
\end{eqnarray}
\end{subequations}
where $\pi_{h0}$, $a_{h1}$, $b_{h0}$ and $b_{h1}$ are independent integration constants. 
As the EOMs for $\pi^{\pm}$, $A_{t}^{\pm}$ and $A_{i}^{\pm}$, i.e., Eqs.~\eqref{eq29d}-\eqref{eq29f}, we can also obtain UV boundary asymptotic expansions as
\begin{subequations}
\label{eq:32}
    \begin{eqnarray}
        \pi^{\pm}(z\rightarrow0)&=&\pi^{\pm}_{0}+\frac{1}{2}\Big\{
   i \varphi_{0}^{\pm}p+i a_{0}^{\pm}(\omega\pm\mu_{I})\nonumber\\
    &&+\pi_{0}^{\pm}(p^{2}-(\omega\pm\mu_{I})^2)
    \Big\}z^2\ln(z)\nonumber\\
    &&+\pi^{\pm}_{2}z^2+\mathcal{O}(z^3),
    \end{eqnarray}
    \begin{eqnarray}
         A_{t}^{\pm}(z\rightarrow0)&=&a_{0}^{\pm}+\frac{1}{2}\Big\{
   a_{0}^{\pm}(p^{2}+g_{5}^2m_{q}^2\zeta^2)\nonumber\\
    &&+\varphi_{0}^{\pm}p\omega+ig_{5}^2m_{q}^2\zeta^2\pi_{0}^{\pm}(\omega\pm\mu_{I})
    \Big\}\nonumber\\
    &&\times z^2\ln(z)+a^{\pm}_{2}z^2+\mathcal{O}(z^3),
    \end{eqnarray}
    \begin{eqnarray}
        A_{i}^{\pm}(z\rightarrow0)&=&\varphi_{0}^{\pm}+\frac{1}{2}\Big\{
    \varphi_{0}^{\pm}g_{5}^2m_{q}^2\zeta^2-ig_{5}^2m_{q}^2p\zeta^2\pi_{0}^{\pm}\nonumber\\
    &&-a_{0}^{\pm}p\omega-\varphi_{0}^{\pm}\omega^2
    \Big\}z^2\ln(z)\nonumber\\
    &&+\varphi_{2}^{\pm}z^2+\mathcal{O}(z^3),
    \end{eqnarray}
\end{subequations}
where $\pi^{\pm}_{0}$, $\pi^{\pm}_{2}$, $a_{0}^{\pm}$, 
 $a_{2}^{\pm}$, $\varphi_{0}^{\pm}$, $\varphi_{2}^{\pm}$ are independent integration constants. Similarly, the horizon asymptotic expansions read
\begin{subequations}
\label{eq:33}
    \begin{eqnarray}
         \pi^{\pm}(z\rightarrow z_{h})&=&(z_{h}-z)^{\frac{i \omega z_{h}}{2\mu_{I}^2z_{h}^2-4}}\Big\{
     \pi_{h0}^{\pm}+\pi_{h1}^{\pm}(z-z_{h})\nonumber\\
    &&+\mathcal{O}[(z-z_{h})^2]
     \Big\}+b_{h0}^{\pm} \nonumber\\
    &&+\frac{(z-z_{h})z_{h}(i b_{h1}^{\pm}z_{h}\pm2b_{h0}^{\pm}\mu_{I})\omega}{16+4z_{h}^4\mu_{I}^4+z_{h}^2(-16\mu_{I}^2+\omega^2)} \nonumber\\
    &&
    +\mathcal{O}[(z-z_{h})^2],\label{eq33a}
    \end{eqnarray}
    \begin{eqnarray}
         A_{t}^{\pm}(z\rightarrow z_{h})&=&(z_{h}-z)^{\frac{i \omega z_{h}}{2\mu_{I}^2z_{h}^2-4}}\Big\{
         a_{h1}^{\pm}(z-z_{h})\nonumber\\
    &&+\mathcal{O}[(z-z_{h})^2]\Big\}-ib_{h0}^{\pm}\omega+b_{h1}^{\pm}(z-z_{h})
     \nonumber\\
    &&+\mathcal{O}[(z-z_{h})^2],\label{eq33b}
    \end{eqnarray}
    \begin{eqnarray}
        A_{i}^{\pm}(z\rightarrow z_{h})&=& i b_{h0}^{\pm}p-\frac{b_{h1}^{\pm}pz_{h}^2\omega(z-z_{h})}{16+4z_{h}^4\mu_{I}^4+z_{h}^2(-16\mu_{I}^2+\omega^2)} \nonumber\\
    &&+\mathcal{O}[(z-z_{h})^2]+(z_{h}-z)^{\frac{i \omega z_{h}}{2\mu_{I}^2z_{h}^2-4}}\nonumber\\
    &&\times\Big\{
         -i\Big[\frac{2 \pi_ {h0}^{\pm} c_ {0}^2 g_ {5}^2(-2 + z_ {h}^2\mu_ {I}^2)}{2pz_{h}^2(-2 + z_ {h}^2\mu_ {I}^2)}\nonumber\\
    &&+\frac{a_ {h1}^{\pm} z_ {h}^2 (4 - 2 z_ {h}^2\mu_ {I}^2 - iz_ {h}\omega)}{2pz_{h}^2(-2 + z_ {h}^2\mu_ {I}^2)}\Big]\nonumber\\
    &&+\varphi_{h1}^{\pm}(z-z_{h})+\mathcal{O}[(z-z_{h})^2]
         \Big\}\label{eq32f}
    \end{eqnarray}
\end{subequations}
with coefficients of first order in Eq.~\eqref{eq33a} and Eq.~\eqref{eq32f},
\begin{subequations}
\begin{eqnarray}
\pi_{h1}^{\pm}&=&\Big\{
-2a_{h1}^{\pm}z_{h}^2(-2+z_{h}^2\mu_{I}^2)^2+\pi_{h0}^{\pm}c_{0}^2(-2+z_{h}^2\mu_{I}^2)\nonumber\\
    &&\times\Big[2g_{5}^2(-2+z_{h}^2\mu_{I}^2)-iz_{h}\lambda\omega\Big]\nonumber\\
    &&+\pi_{h0}^{\pm}z_{h}^2\Big\{2p^2(-2+z_{h}^2\mu_{I}^2)^2+i\omega\Big[4z_{h}^5\mu_{g}^2\mu_{I}^4\nonumber\\
    &&+z_{h}^3\mu_{I}^2(2\mu_{c}^2-16\mu_{g}^2-3\mu_{I}^2)+z_{h}(-4\mu_{c}^2+16\mu_{g}^2\nonumber\\
    &&+6\mu_{I}^2)+2i(4\mu_{I}+3\omega)-iz_{h}^2\mu_{I}^2(4\mu_{I}+9\omega) \Big] \Big\}
\Big\}\nonumber\\
    &&  
   \Big/\Big[4z_{h}(-2+z_{h}^2\mu_{I}^2)^2(-2+z_{h}^2\mu_{I}^2+iz_{h}\omega)
\Big],
\end{eqnarray} 
\begin{eqnarray}
    \varphi_{h1}^{\pm}&=&i\Big\{
-4\pi_{h0}^{\pm}c_{0}^2g_{5}^2p^2z_{h}^2(-2+z_{h}^2\mu_{I}^2)^3\nonumber\\&&+2ia_{h1}^{\pm}p^2z_{h}^5(-2+z_{h}^2\mu_{I}^2)^2\omega+\Big[-2\pi_{h0}^{\pm}c_{0}^2g_{5}^2(-2\nonumber\\&&+z_{h}^2\mu_{I}^2)+a_{h1}^{\pm}z_{h}^2(-4+2z_{h}^2\mu_{I}^2+iz_{h}\omega)
    \Big]\nonumber\\&&\times\Big[2c_{0}^2g_{5}^2(-2+z_{h}^2\mu_{I}^2)^2+iz_{h}\omega\Big(-4+4z_{h}^6\mu_{g}^2\mu_{I}^4\nonumber\\&&+16z_{h}^2(\mu_{g}^2+\mu_{I}^2)-z_{h}^4(16\mu_{g}^2\mu_{I}^2+7\mu_{I}^4)\nonumber\\&&+6iz_{h}\omega-9iz_{h}^3\mu_{I}^2\omega
    \Big)
    \Big]
    \Big\}\Big/\Big\{
8z_{h}^3(-2+z_{h}^3\mu_{I}^2)^4\nonumber\\&&\times\Big(p+\frac{ipz_{h}\omega}{-2+z_{h}^2\mu_{I}^2}
    \Big)
    \Big\},
\end{eqnarray}
\end{subequations}
where the independent integration constants are $\pi_{h0}^{\pm}$, $a_{h1}^{\pm}$, $b_{h0}^{\pm}$ and  $b^{\pm}_{h1}$. The on-shell action of pion part is
\begin{eqnarray}
    S_{\pi_{}}^{on}&=&\frac{1}{2g_{5}^2}\int d^4k \sum_{a=1}^3
    \Big\{
   e^{A-\Phi}\Big[A_{t}^a(-k,z)\partial_{z}A_{t}^a(k,z)\nonumber\\
   &&-fA_{i}^a(-k,z)\partial_{z}A_{i}^a(k,z)
   \Big]\nonumber\\
   &&-e^{3A-\Phi}g_{5}^2f\chi^2\pi^a(-k,z)\partial_{z}\pi^a(k,z)
    \Big\}\Big|_{z=\epsilon}^{z=z_{h}},
 \label{eq33} 
\end{eqnarray}
with $k$ the four dimension momentum. Substituting the Eqs.~\eqref{eq30}-\eqref{eq:33} into Eq.~\eqref{eq33} and taking derivative with respect to the source $J_{\pi}$, we can obtain the on-shell action and the retarded correlator of $\pi^0$ as follow
\begin{figure*}[bhtp]
\centering
    \begin{overpic}[width=0.42\textwidth]{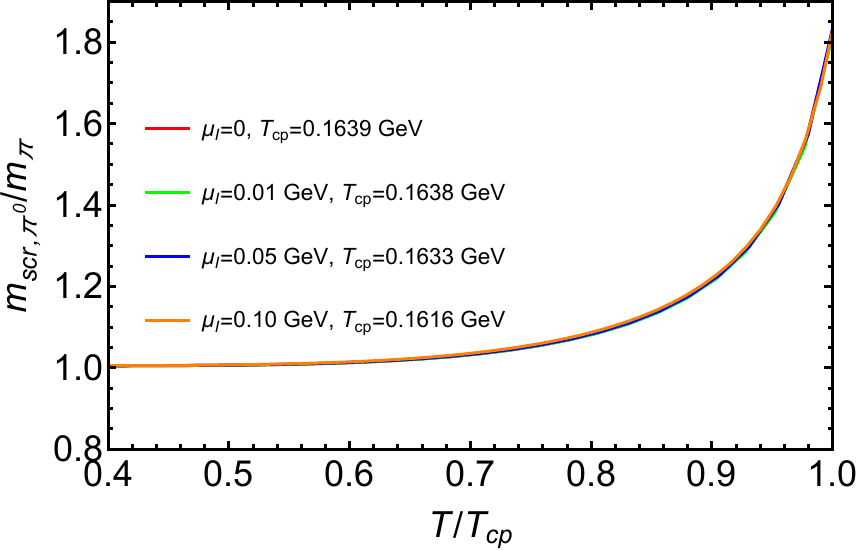}
        \put(18,55){\bf{(a)}} 
        \label{fig3a}
    \end{overpic}
    \begin{overpic}[width=0.42\textwidth]{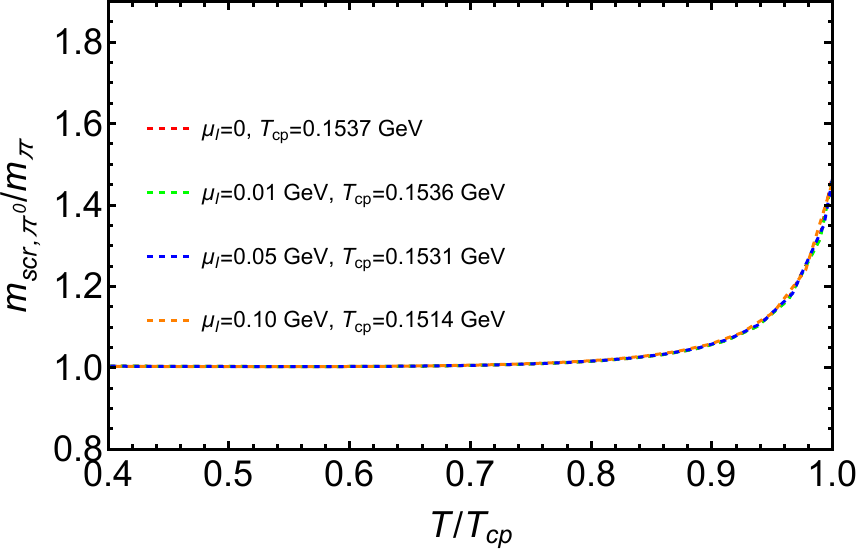}
        \put(18,55){\bf{(b)}} 
        \label{fig3b}
    \end{overpic}
    \begin{overpic}[width=0.42\textwidth]{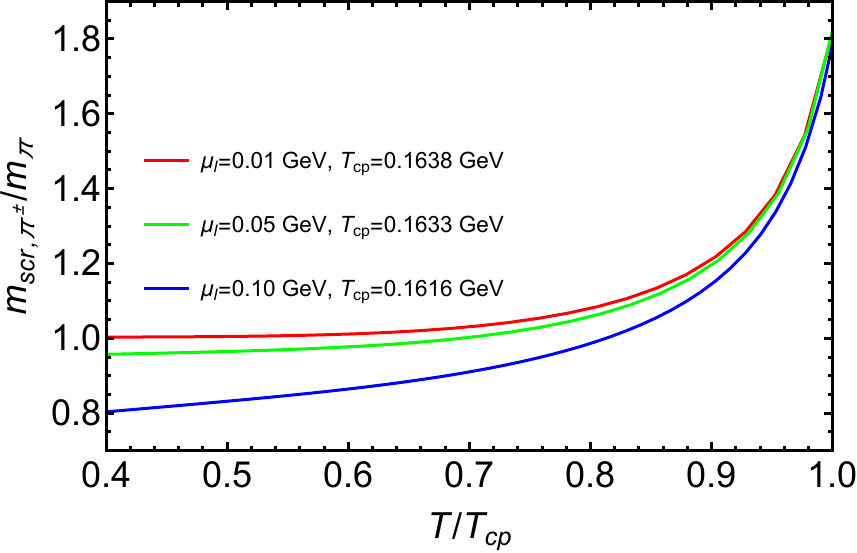}
        \put(18,55){\bf{(c)}} 
        \label{fig3c}
    \end{overpic}
     \begin{overpic}[width=0.42\textwidth]{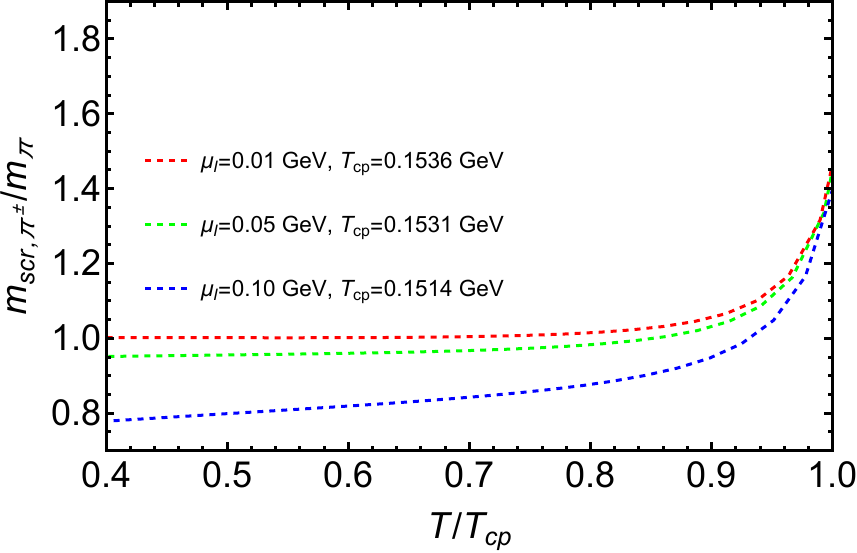}
        \put(18,55){\bf{(d)}} 
        \label{fig3d}
    \end{overpic}
    \caption{Screening mass as functions of $T$ of $\pi^0$ in \textbf{(a)} model I and \textbf{(b)} model II. The red, green, blue, and orange solid lines stand for the results below $T_{cp}$, at $\mu_{I}=0, 0.01, 0.05, 0.10$ GeV, respectively. Screening mass as functions of $T$ of charged pions  $\pi^{\pm}$ in \textbf{(c)} model I and \textbf{(d)} model II. The red, green,and blue solid lines stand for the results below $T_{cp}$, at $\mu_{I}=0.01, 0.05, 0.10$ GeV, respectively.
    }
    \label{fig3}
\end{figure*}

\begin{eqnarray}
    S_{\pi^0}^{on}&=&\frac{1}{g_{5}^2}\int d^4k\Big\{
    a_{0}(-k)a_{2}(k)-\varphi_{0}(-k)\varphi_{2}(k)\nonumber\\
    &&-g_{5}^2m_{q}^2\zeta^2\pi_{0}(-k)\pi_{2}(k)
    \Big\},\\
    G_{\pi^{0}}(k)&=&\frac{\delta^2 S_{\pi^0}^{on}}{\delta J_{\pi^0}^{*}\delta J_{\pi^0}}=-m_{q}^2\zeta^2\frac{\pi_{2}(k)}{\pi_{0}(k)}.\label{eq35}
\end{eqnarray}
The the on-shell actions and the retarded correlator of $\pi^{\pm}$ read

\begin{eqnarray}
        S_{\pi^{\pm}}^{on}&=&\frac{1}{g_{5}^2}\int d^4k
    \Big\{
     a_{0}^{\pm}(k)^*a_{2}^{\pm}(k)-\varphi_{0}^{\pm}(k)^*\varphi_{2}^{\pm}(k)\nonumber\\
    &&-g_{5}^2m_{q}^2\zeta^2\pi_{0}^{\pm}(k)^*\pi_{2}^{\pm}(k)
    \Big\},\\
     G_{\pi^{\pm}}(k)&=&\frac{\delta^2 S_{\pi^{\pm}}^{on}}{\delta J_{\pi^{\pm}}^{*}\delta J_{\pi^{\pm}}}=-m_{q}^2\zeta^2\frac{\pi_{2}^{\pm}(k)}{\pi_{0}^{\pm}(k)}.
     \label{eq37}
\end{eqnarray}

\subsection{\label{subsecb} Screening masses of pions}
In this section, we will solve the EOMs of pions numerically and exact the screening masses of the neutral pion and the charged pions from the pole of the spatial correlation function. Then we will investigate the temperature as well as the isospin chemical potential dependent behavior of screening mass with finite physical quark mass. In this paper, we will mainly focus on the temperature region below $T_{cp}$. Not only does the pion condensate occur below the chiral transition temperature, but also the particles of pions are not well-defined degrees of freedom at high temperature.

From~Eq.~\eqref{eq19}, the screening mass is extracted from the pole of the retarded correlator when the frequency $\omega=0$. For the retarded correlator of neutral pion in ~Eq.~\eqref{eq35}, it means the vanishing UV boundary conditions,
\begin{equation}
\label{eq40}
    \pi_{0}({p}^2)=0,\quad a_{0}({p}^2)=0,\quad \varphi_{0}({p}^2)=0.
\end{equation}
The lowest state of ${p}^2$ satisfying Eq.~\eqref{eq40} corresponds to screening mass of the neutral pion, i.e., $m_{\rm{scr}}^2=-{p}^2$.
 What is more, since the equations for $\pi^0$, $A_{t}^0$ and $A_{i}^0$ (Eqs.~\eqref{eq29a}-\eqref{eq29c}) are linear differential equations, we can set the integrate constant $\pi_{h0}$ at the horizon to be unity ($\pi_{h0}=1$) without shifting the mass spectra. The integration constant $b_{h1}$ should be set to zero ($b_{h1}=0$) which insure the on-shell action $S_{\pi}^{on}$ is independent of horizon terms~\cite{Son:2002sd}. Finally, the remain undetermined integration constants, $b_{h0}$ and $a_{h1}$, can be also determined by ``shooting method''~\cite{boyd2001chebyshev}. The same prescription can be also applied to the EOMs of charged pions in Eqs.~\eqref{eq29d}-\eqref{eq29f}. At the horizon, we can take $\pi_{h0}^{\pm}=1$, while $b_{h0}^{\pm}$, $a_{h1}^{\pm}$ and ${p}^2$ can be determined by ``shooting method'' when the following conditions at the UV boundary are satisfied,
\begin{equation}
    \pi_{0}^{\pm}({p}^2)=0,\quad a_{0}^{\pm}({p}^2)=0,\quad \varphi_{0}^{\pm}({p}^2)=0,
\end{equation}
simultaneously.
They induce to the pole of the correlator in Eq.~\eqref{eq37}.
 However, there are some differences related to integration constant $b_{h1}^{\pm}$ at the horizon at finite isospin chemical potential. We find that only the condition  $\partial_{z}\pi^{\pm}(z\rightarrow z_{h})=0$, i.e., $b_{h1}^{\pm}=\pm2ib_{h0}^{\pm}\mu_{I}/z_{h}$ predicts appropriate spectra of charged pions. When $\mu_{I}=0$, $b_{h1}^{\pm}$ equals zero, which is consistent with the previous discussions.
 Under these boundary conditions mentioned above, we can solve the equations of motion and extract the screening masses of neutral pion $\pi^0$ as well as charged pions $\pi^{\pm}$.
\begin{figure*}
    \centering
     \begin{overpic}[width=0.42\textwidth]{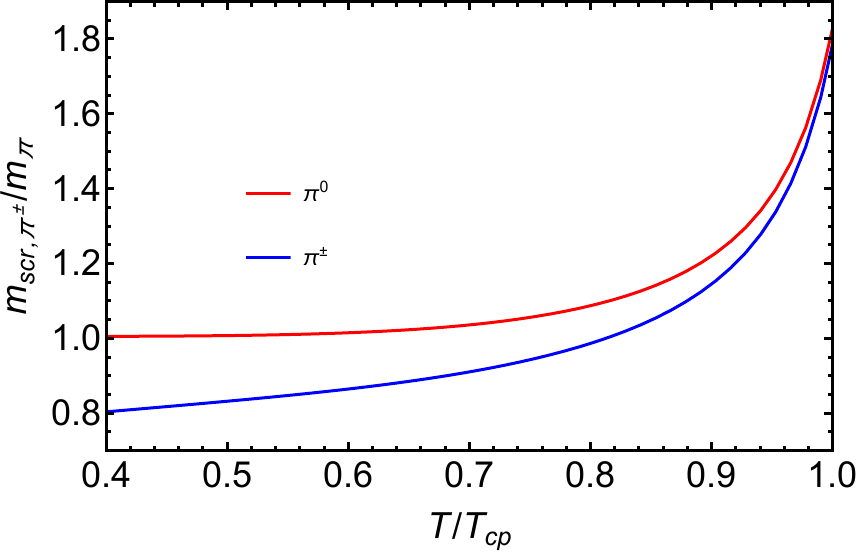}
        \put(18,55){\bf{(a)}} 
        \label{fig3e}
    \end{overpic}
     \begin{overpic}[width=0.42\textwidth]{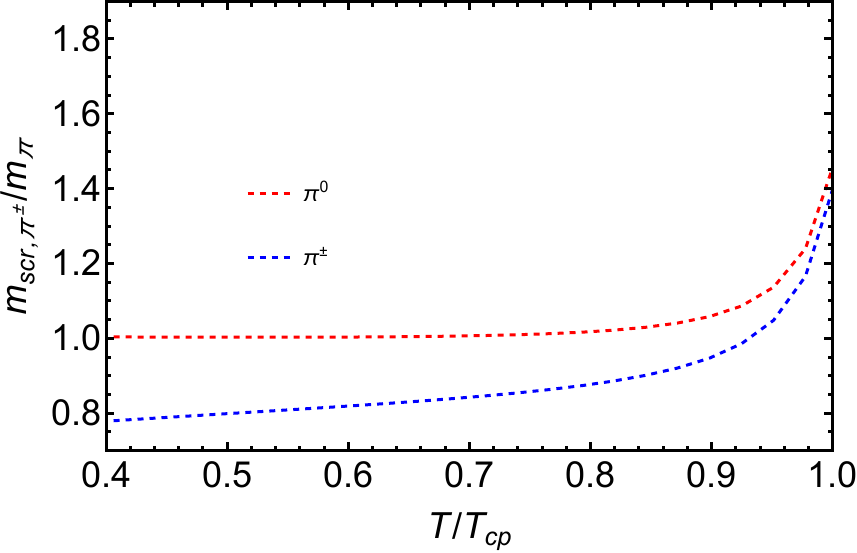}
        \put(18,55){\bf{(b)}} 
        \label{fig3f}
    \end{overpic}
    \caption{Screening mass as functions of $T$ of $\pi^0$ as well as $\pi^{\pm}$ in \textbf{(a)} model I and \textbf{(b)} model II at fixed isospin chemical potential $\mu_{I}=0.10$ GeV.}
    \label{fig3p}
\end{figure*}

\subsubsection{The temperature effect}
The numerical results of screening masses, varying with temperature,  are presented in Fig.~\ref{fig3}. Qualitatively, one can find that these two soft-wall AdS/QCD models share the same features in predicting the temperature behavior of screening masses of pions. For the screening mass of $\pi^0$, the numerical results at different $\mu_{I}$ as functions of $T$ in model I and model II, are shown in Figs.~\ref{fig3}(a) and (b), respectively. From which we can find that the screening mass of neutral pion at very low temperature is almost independent on $T$. The value of $m_{\rm{scr},\pi^0}/m_{\pi}$ \footnote{Here, $m_{\pi}\approx0.13971136$ GeV, which is the pion mass at $T=0$, $\mu_{I}=0$ in model I~\cite{Fang:2016nfj}, and $m_{\pi}\approx0.13971648$ GeV in model II.} gets closed to 1, which implies that the screening mass of $\pi^0$ at low temperature almost remains $m_{\pi}$. As the rising of $T$, $m_{\rm{scr},\pi^0}$ grows slowly first and then enhances quickly when $T$ is close to the pseudo-critical temperature $T_{cp}(\mu_{I})$~\footnote{From the discussion in Section~\ref{sec2}, it can be seen that $T_{cp}$ is affected by $\mu_{I}$. It is found that $T_{cp}(\mu_{I})$ will decrease as the rising of $\mu_{I}$.}. What is more, it is noteworthy that $m_{\rm{scr},\pi^0}$ at different $\mu_{I}$ have almost the same value. It is because that $\pi^0$ is isospin chargeless and the isospin chemical potential has little impact on $m_{\rm{scr},\pi^0}$. 

For the screening masses of charged pions $\pi^{\pm}$, the numerical results at different $\mu_{I}$ are shown in Figs.~\ref{fig3}(c) and (d). As a result of  finite isospin chemical potential, we find that the screening masses of $\pi^+$ and $\pi^-$ are degenerate, which is agreement with the NJL model results in Ref.~\cite{Jiang:2011aw}. With the increasing of temperature, the screening masses increase. However, with the increasing of isospin chemical potential, the screening masses decrease. For example, we can see in both models that at fixed temperature $T/T_{cp}(\mu_{I})=0.4$, 
\begin{eqnarray}
m_{\rm{scr},\pi^{\pm}}(\mu_{I}=0.01~  {\rm GeV})& \approx & m_{\pi}, \nonumber\\
 m_{\rm{scr},\pi^{\pm}}(\mu_{I}=0.05~  {\rm GeV})& \approx & 0.95 m_{\pi}, \nonumber\\
  m_{\rm{scr},\pi^{\pm}}(\mu_{I}=0.10~  {\rm GeV})& \approx &0.8 m_{\pi}.\nonumber
\end{eqnarray}
 Furthermore, in the high temperature region, the effect of $\mu_{I}$ is much weaker than the temperature effect. The curves become degenerate when $T$ closes to $T_{cp}$.
\begin{figure}[htbp]
\centering
    \begin{overpic}[width=0.42\textwidth]{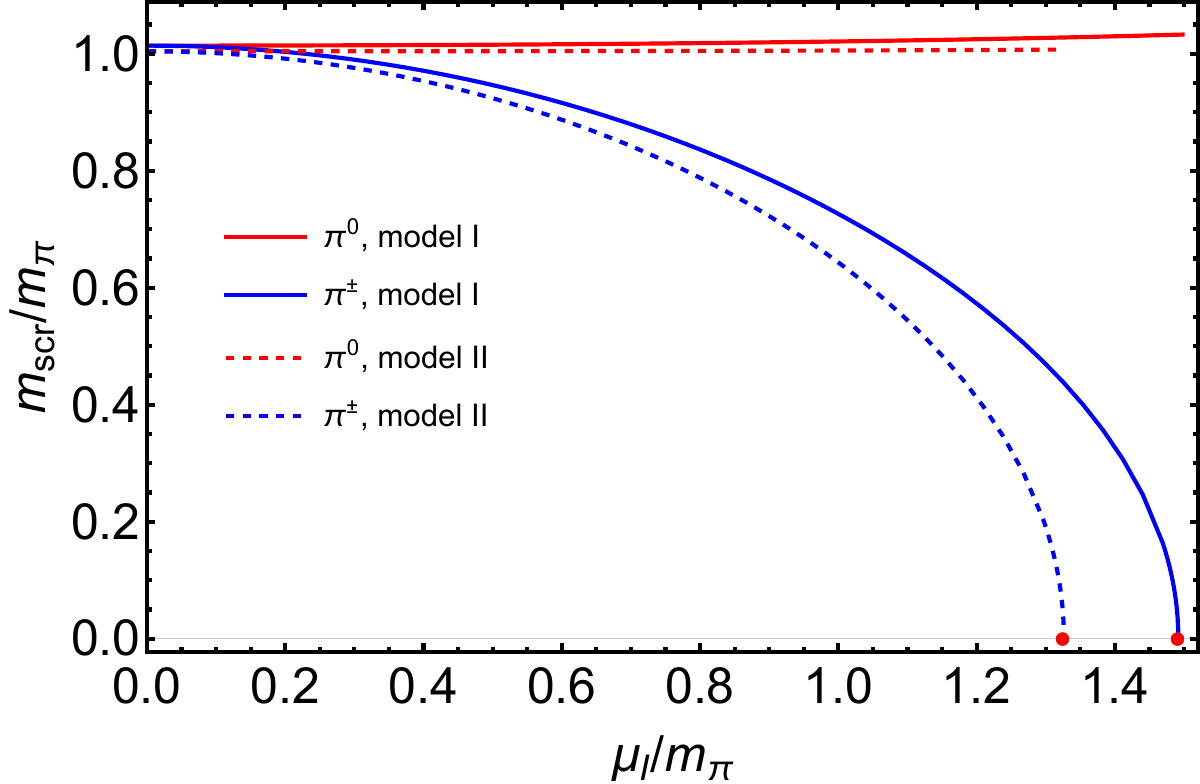}
    \end{overpic}
    \caption{Screening mass as functions of isospin chemical potential $\mu_{I}$ at fixed temperature $T=0.10$ GeV in model I (solid lines) and model II (dashed lines). The red and blue solid lines stand for $\pi^0$ and $\pi^{\pm}$, respectively. Screening mass $m_{\rm{scr}}$ vanishes at 1.49 $m_{\pi}\approx 0.208$ GeV in model I and 1.32 $m_{\pi}\approx 0.184$ GeV in model II, as shown by the red dots.}
     \label{fig4}
\end{figure}

To compare the screening masses between the neutral and charged pions, we show the temperature dependent curves of the screening masses at fixed isospin chemical potential $\mu_{I}=0.10$ GeV in Figs.~\ref{fig3p}. In both models, the screening mass of $\pi^0$ gets closed to $m_{\pi}$ while $\pi^{\pm}$ is about $0.8 m_{\pi}$ at low temperature. However, they become degenerate at relatively high temperatures, which implies the neglectable effects of $\mu_{I}$.

\subsubsection{The isospin density effect}

After considering the temperature dependence of the screening masses at fixed isospin chemical potential, we will discuss the effect of isospin chemical potential on screening masses at fixed temperatures in this subsection. Due to the qualitative consistency of the conclusions obtained at different temperatures, we only choose the fixed temperature $T=0.10$ GeV for the discussion without loss of generality. The numerical results of the isospin chemical potential dependence of the screening masses in both models are presented in Fig.~\ref{fig4}. Both of these models exhibit the same behaviors. In the normal phase, i.e., $\mu_{I}<\mu_{I}^c$, $m_{\rm{scr},\pi^0}$ and $m_{\rm{scr},\pi^{\pm}}$ are splitting, since the EOMs of charged pions, Eqs.~\eqref{eq29}(a)-(c), depend on isospin density through $\nu(z)$. However, the EOM of neutral pion, Eqs.~\eqref{eq28}(a)-(c), does not.
The neutral pion $\pi^0$ almost keeps unchanged with the increasing of isospin chemical potential. This is consistent with the previous discussion since $\pi^0$ does not carry isospin charge. What is more, $m_{\rm{scr},\pi^{\pm}}$ decrease monotonically with $\mu_{I}$ and vanish at a critical chemical potential $\mu_{I}^c$, where the pion superfluid phase transition occurs. For the chosen temperature $T=0.10$ GeV, the critical isospin chemical potential $\mu_{I}^c$ in model I is about $1.49 m_{\pi}\approx 0.208$ GeV. For model II, it is $1.31 m_{\pi}\approx 0.183 ~\rm{GeV}$. When the chemical potential is beyond $\mu_{I}^c$, the pion condensation will happen. The discussions of pions' properties in the pion superfluid phase will be left to our future works.

Qualitatively, both of these two different soft-wall AdS/QCD models possess the same behaviors of the screening masses. Furthermore, these behaviors are well consistent with the results of the NJL model in Ref.~\cite{Jiang:2011aw}.

\subsection{Pole masses of pions}
\begin{figure*}[htbp]
\centering
    \begin{overpic}[width=0.42\textwidth]{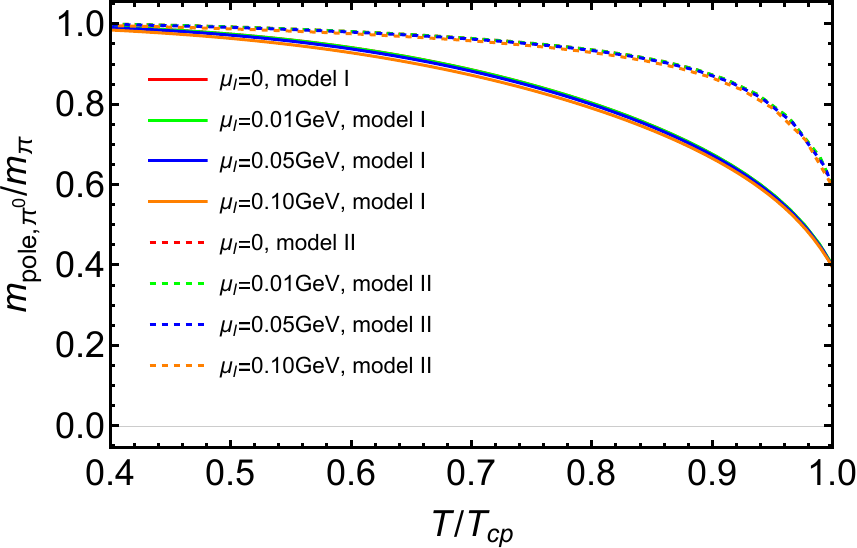}
        \put(88,55){\bf{(a)}} 
    \end{overpic}
    \begin{overpic}[width=0.42\textwidth]{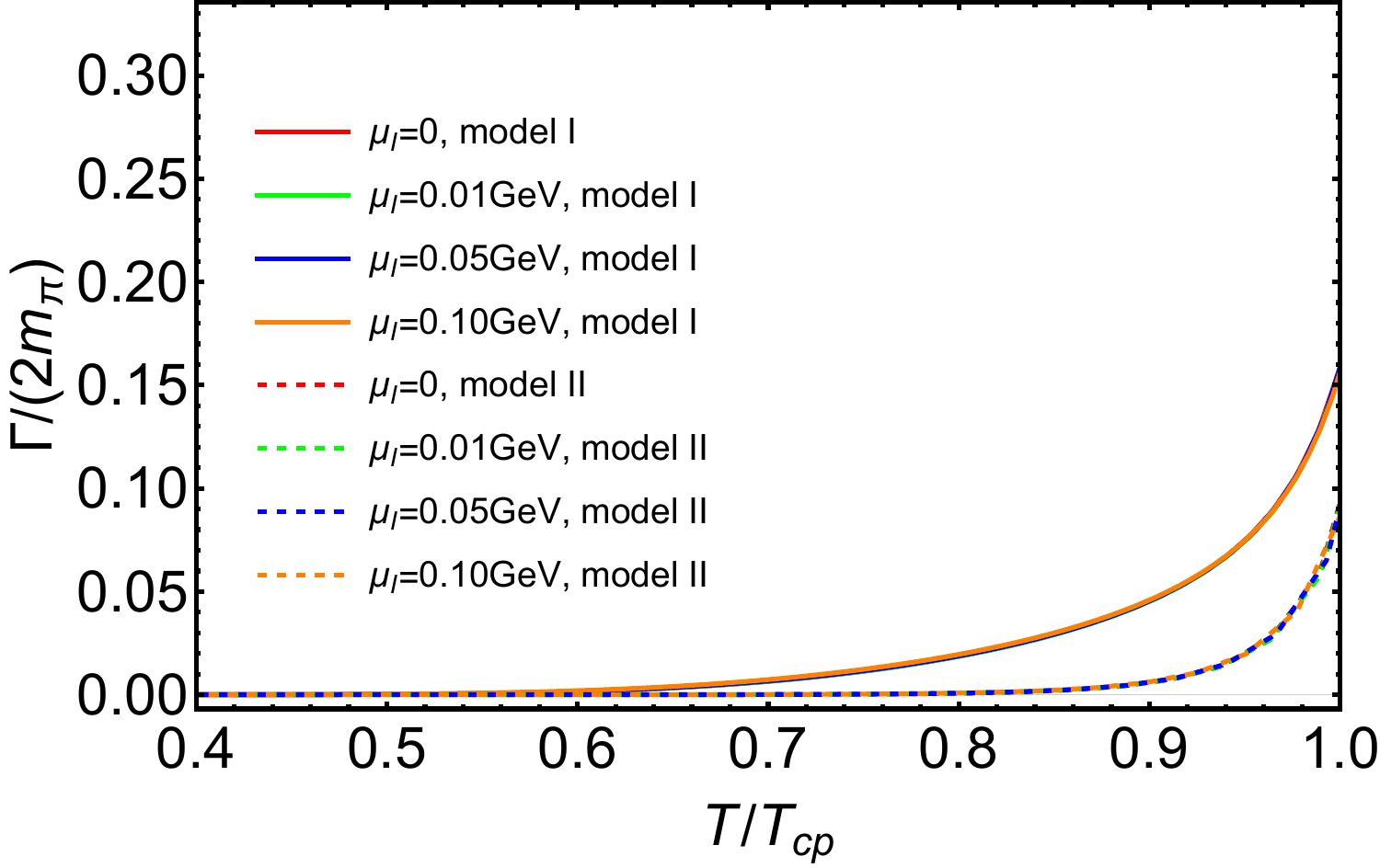}
        \put(88,55){\bf{(b)}} 
    \end{overpic}
    \begin{overpic}[width=0.42\textwidth]{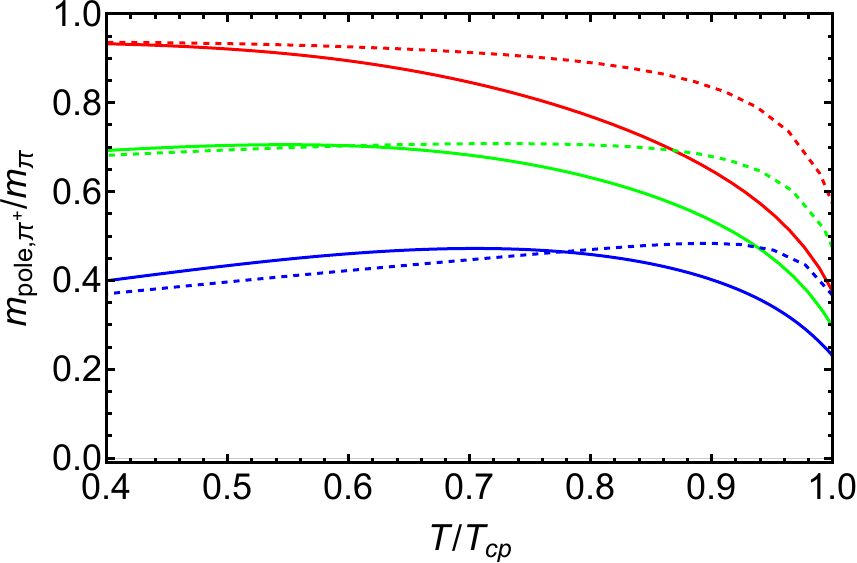}
        \put(88,55){\bf{(c)}} 
    \end{overpic}
     \begin{overpic}[width=0.42\textwidth]{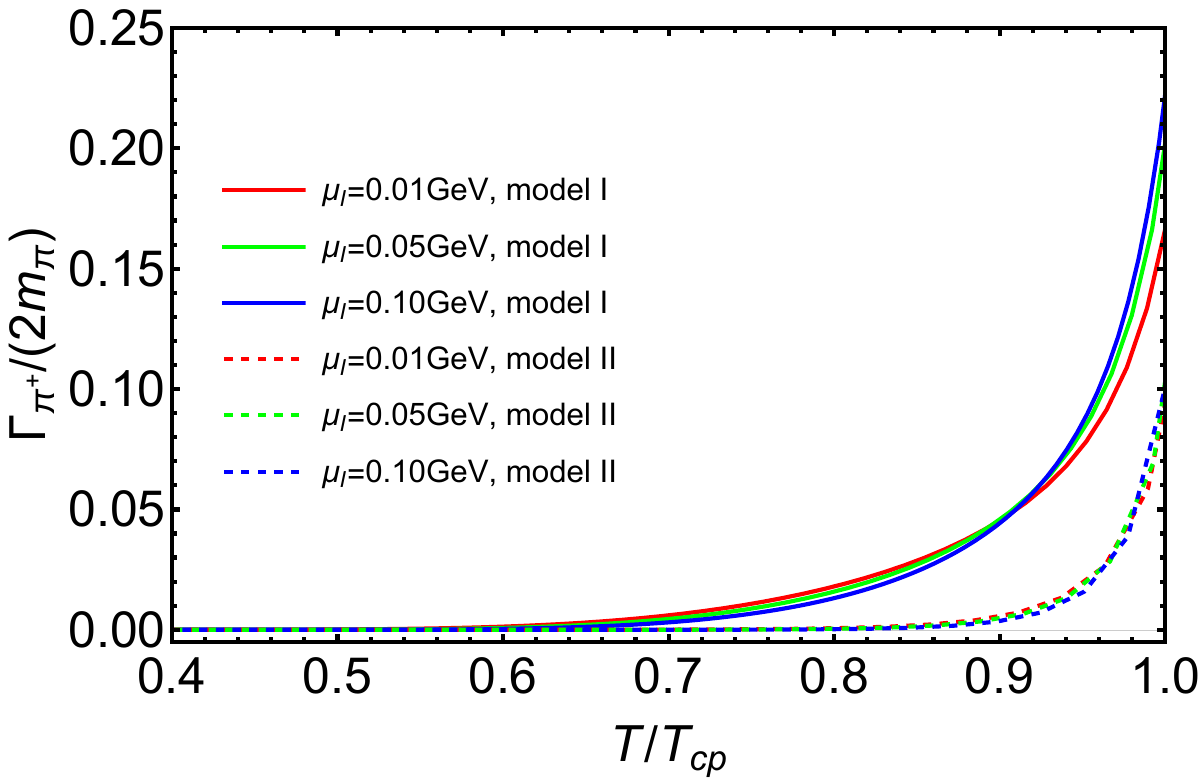}
        \put(88,55){\bf{(d)}} 
    \end{overpic}
     \begin{overpic}[width=0.42\textwidth]{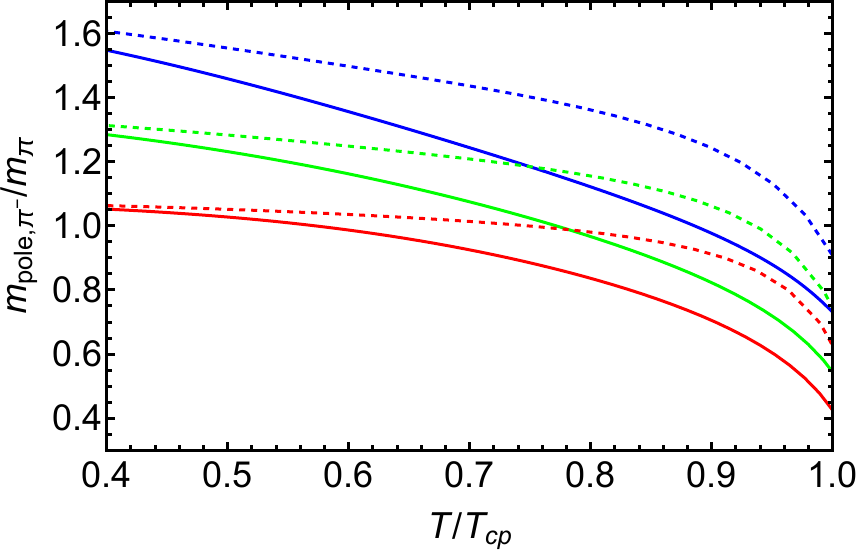}
        \put(88,55){\bf{(e)}} 
    \end{overpic}
     \begin{overpic}[width=0.42\textwidth]{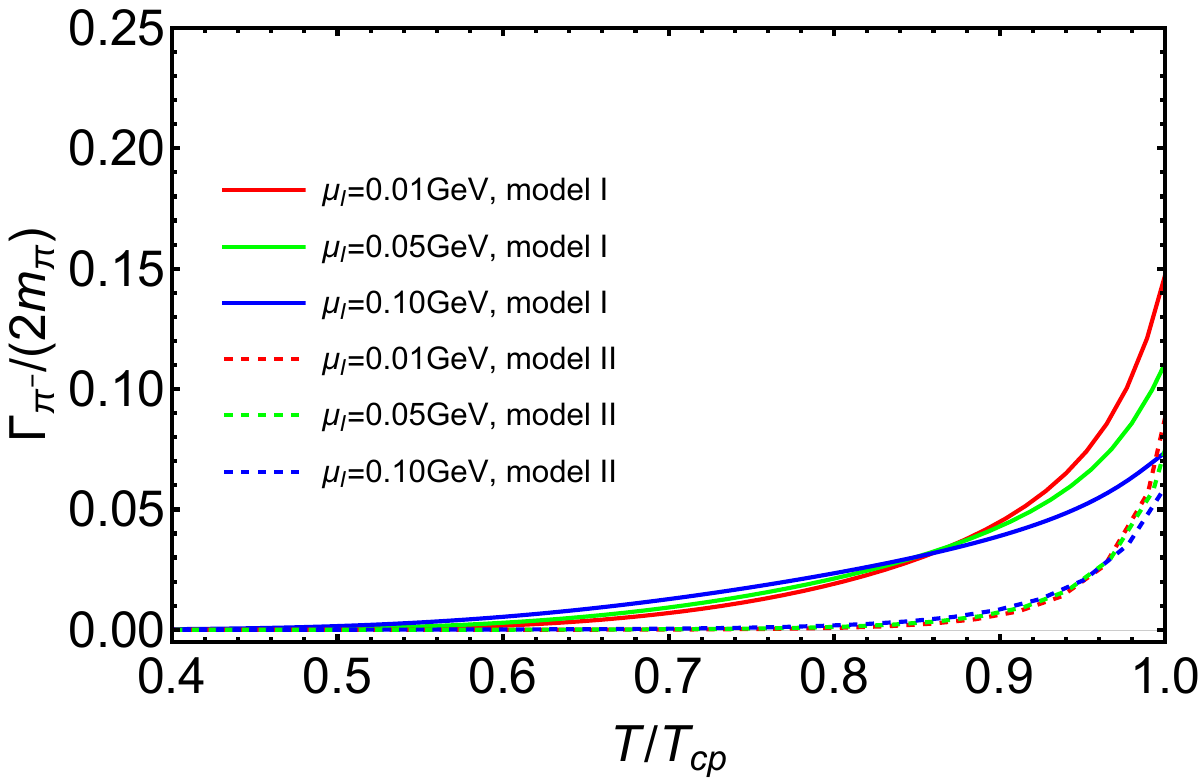}
        \put(88,55){\bf{(f)}} 
    \end{overpic}
    \caption{Pole mass and thermal width as functions of $T$ for \textbf{(a)}, \textbf{(b)} $\pi^0$; \textbf{(c)}, \textbf{(d)} $\pi^+$ and \textbf{(e)}, \textbf{(f)} $\pi^-$. In (a) and (b), the red, green, blue and orange lines stand for results at $\mu_{I}=0$, $ 0.01$, $0.05$, and $0.10$ GeV, respectively. In (c),(d),(e) and (f), the red, green and blue lines represent results at $\mu_{I}=0.01$, 0.05 and 
 0.10 GeV, respectively. The solid lines represent results in model I while the dashed lines represent results in model II.
    }
    \label{fig5}
\end{figure*}

In the framework of holographic approach, one can obtain the pole masses of pions from the peak positions of spectral functions which are related to the imaginary part of the two-point retarded correlation functions. However, as pointed out in Refs.~\cite{Cao:2020ryx,Cao:2021tcr} that the thermal widths of quasiparticle pions will also increase with the increasing temperature, which leads to inconspicuous resonance peaks of the spectral functions. A more straightforward and effective approach to define the effective masses of the pions is through the corresponding QNM. The quasi-normal frequency is $\omega_{0}$, which corresponds to the pole of the temporal part of the retarded correlator $G(\omega)$. Its real part and imaginary one correspond to the meson's pole mass $m_{\rm{pole}}=\rm{Re}(\omega_0)$ and thermal width $\Gamma=-2\rm{Im}(\omega_0)$, respectively. 

We will numerically calculate the quasi-normal frequency $\omega_0$ with two different soft-wall AdS/QCD models. Then, we directly obtain the pole masses $m_{\rm{pole}}$ and the thermal widths $\Gamma$ from $\omega_0$. For the pole mass and thermal width, we need only focus on the temporal part of the correlator in Eqs.~\eqref{eq35} and \eqref{eq37}, so that we take ${p}=0$ in the corresponding EOMs in Eqs.~\eqref{eq29} and~\eqref{eq28}, and the boundary conditions in Eqs.~\eqref{eq30}-\eqref{eq:33}. Note that the equations of $A_{i}^a$ are decoupled from $\pi^a$ and $A_{t}^a$ at ${p}=0$. Therefore, we only need to solve the equations for $\pi^a$ and $A_{t}^a$ in Eqs.~\eqref{eq29a},~\eqref{eq29b} and~\eqref{eq29d},~\eqref{eq29e}. In order to determine the particular QNM frequency $\omega=\omega_0$, which corresponds to the pole of the retarded correlator,  the integration constants at the UV boundary should hold the following relations,
\begin{equation}
    \pi_{0}(\omega=\omega_0)=a_{0}(\omega=\omega_0)=0
\end{equation}
for the neutral pion, and 
\begin{equation}
    \pi_{0}^{\pm}(\omega=\omega_0)=a_{0}^{\pm}(\omega=\omega_0)=0
\end{equation}
for the charged pions.
On the horizon, similar to our previous discussions for the screening mass in the Section.~\ref{subsecb}, one can set 
\begin{equation}
\pi_{h0}=1,\ \ \ b_{h1}=0
\end{equation}
for the neutral pion, and
\begin{equation}
\pi_{h0}^{\pm}=1,\ \ \ b_{h1}^{\pm}=\pm 2ib_{h0}^{\pm}\mu_I/z_h
\end{equation}
for the charged pions. With these boundary conditions, one can numerically solve the EOMs for the neutral pion and charged pions and obtain the QNM frequency with the ``Shooting method''.

\begin{figure*}[htbp]
\centering
    \begin{overpic}[width=0.42\textwidth]{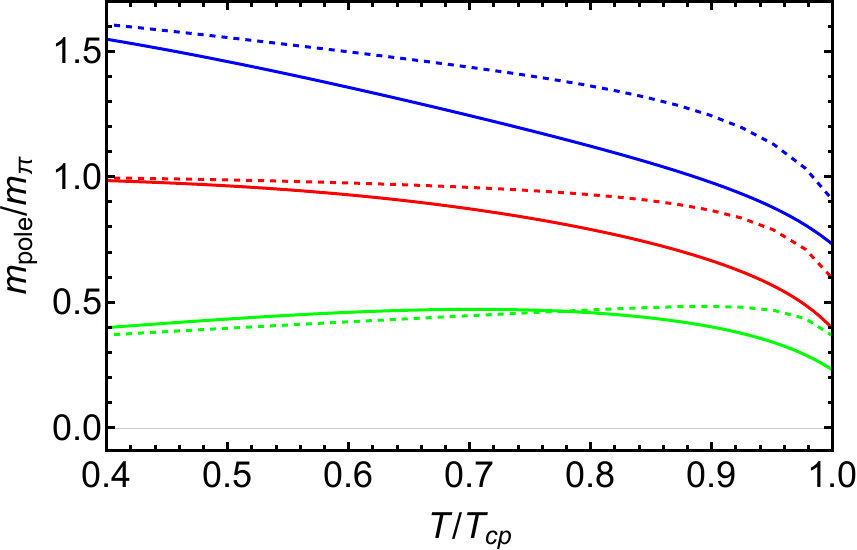}
        \put(88,55){\bf{(a)}} 
    \end{overpic}
    \begin{overpic}[width=0.42\textwidth]{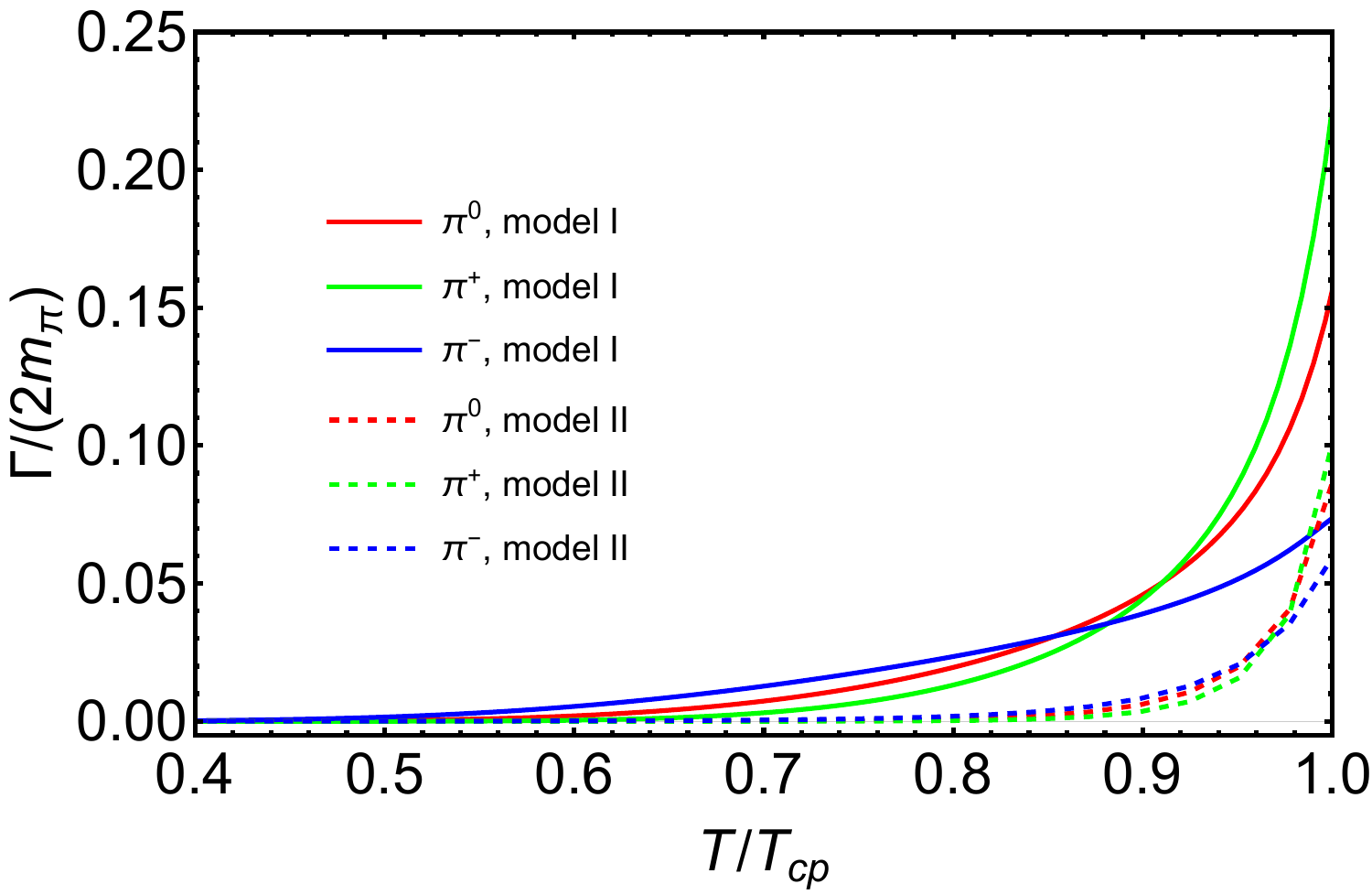}
        \put(88,55){\bf{(b)}} 
    \end{overpic}
    \caption{\textbf{(a)} Pole masses and \textbf{(b)} thermal widths as function of $T$ at $\mu_{I}=0.10$ GeV. The blue, red and green lines represent results for $\pi^-, \pi^0, \pi^+$, respectively. The solid lines stand for results in model I and dashed lines stand for results in model II.
    }
    \label{fig6}
\end{figure*}

\subsubsection{The temperature effect}
To investigate the temperature dependence of pole masses and thermal width of $\pi^0$, $\pi^+$ and $\pi^-$ in both models, we fix isospin chemical potential and vary temperature. As the same reasons illustrated in Sec.~\ref{subsecb}, we just pay close attention to the pole masses of pions in the normal phase at the temperature $T<T_{cp}$. We consider the cases with fixed isospin chemical potential at $\mu_{I}=0$, $0.01$, $0.05$, and $0.10$ GeV, respectively. The corresponding numerical results are shown in Fig.~\ref{fig5}.
In Fig.~\ref{fig5}(a), the pole mass $m_{\rm{pole}}$ decreases monotonously with the increasing of temperature. We have 
\begin{eqnarray}
&{{\rm Model\  I}}&\quad m_{\rm{pole},\pi^0}(T=T_{cp})\approx 40\% m_\pi ,\nonumber\\
&{{\rm Model\  II}}&\quad m_{\rm{pole},\pi^0}(T=T_{cp})\approx 70\% m_\pi ,\nonumber
\end{eqnarray}
where $m_\pi$ is the model dependent pole mass with $\mu_I=0$ and $T=0$. Qualitatively, the decreasing behavior around the pseudo-critical temperature in both soft-wall AdS/QCD models, are consistent with T. D. Son et al's analytical analysis through the chiral perturbation theory in Refs.~\cite{Son:2002ci,Son:2001ff}. In Fig.~\ref{fig5}(b), the thermal width $\Gamma$ increases monotonously with the increasing of temperature. One can see that the results for $\pi^0$ are almost not affected by $\mu_I$, because it does not carry isospin charge.

The results for $\pi^+$ at different fixed isospin chemical potential, are shown in Figs.~\ref{fig5}(c) and (d). From Fig.~\ref{fig5}(c), we find that $\mu_{I}$ depress the pole mass of $\pi^+$, i.e. the larger $\mu_I$ the the lower $m_{\rm{pole},\pi^+}$. For example, at $T/T_{cp}=0.4$, the reduction of $m_{\rm{pole},\pi^+}$ at $\mu_{I}=0.05$ GeV and 0.10 GeV are about 30\% and 60\%, respectively. This result may be reasonable since one might expect the gathering of positive isospin charge would make it easier to excite a $\pi^+$. When $\mu_{I}$ is small, such as 
$\mu_{I}=0.01$ GeV, $m_{\rm{pole},\pi^+}$ decreases monotonously with the increasing of temperature. However, when $\mu_{I}$ getting larger, such as $\mu_{I}=0.10$ GeV, $m_{\rm{pole},\pi^+}$ increases first at low temperature and decreases when temperature is close to $T_{cp}(\mu_{I})$.  
In Fig.~\ref{fig5}(d), the thermal widths of $\pi^+$ at different fixed $\mu_{I}$ increase very slowly and the effect of $\mu_{I}$ is not obvious at low temperature. When $T$ gets closed to $T_{cp}$, it increases quickly. 

As to the negative charged pions $\pi^-$, the numerical results of the pole masses $m_{\rm{pole},\pi^-}$ and thermal widths $\Gamma_{\pi^-}$ are presented in Figs.~\ref{fig5}(e) and~(f). From the results we can find that $m_{\rm{pole},\pi^-}$ decreases monotonously with the increasing of temperature. At a fixed temperature, $m_{\rm{pole},\pi^-}$ is enhanced by $\mu_{I}$. For example, at $T/T_{cp}=0.4$ GeV and $\mu_{I}=0.10$ GeV, $m_{\rm{pole},\pi^-}$ is increased by about 55\% $m_\pi$ in model I and by about 60\% $m_\pi$ in model II, respectively. While the thermal width increases with the increasing of temperature. 

To compare the different isospin effects among the neutral and charged pions, we show the temperature-dependent pole mass and thermal width of $\pi^0, \pi^+, \pi^-$  at fixed $\mu_{I}=0.10$ GeV in Fig.~\ref{fig6}, simultaneously. As a result of finite $\mu_{I}$, the ${\rm{SU}}_{\rm I}(2)$ symmetry is explicit broken and the pions split, which is obviously shown in Fig.~\ref{fig6}, $m_{\rm{pole},\pi^-}>m_{\rm{pole},\pi^0}>m_{\rm{pole},\pi^+}$. As the temperature approaches $T_{cp}$, the pole masses of pions decrease and tend to degeneracy due to the restoration of the chiral symmetry. The thermal widths of neutral and charged pions increase with the increasing of temperature.

\begin{figure*}[htb]
\centering
    \begin{overpic}[width=0.40\textwidth]{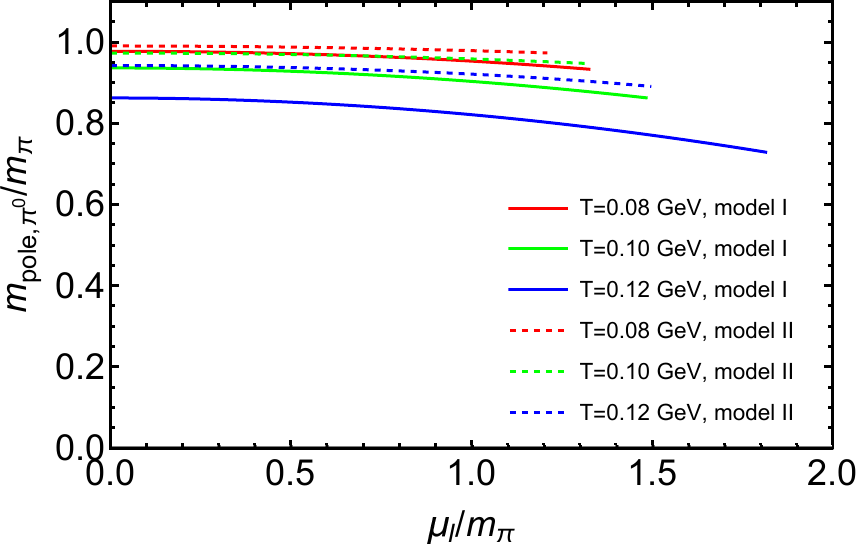}
        \put(88,55){\bf{(a)}} 
    \end{overpic}
    \begin{overpic}[width=0.42\textwidth]{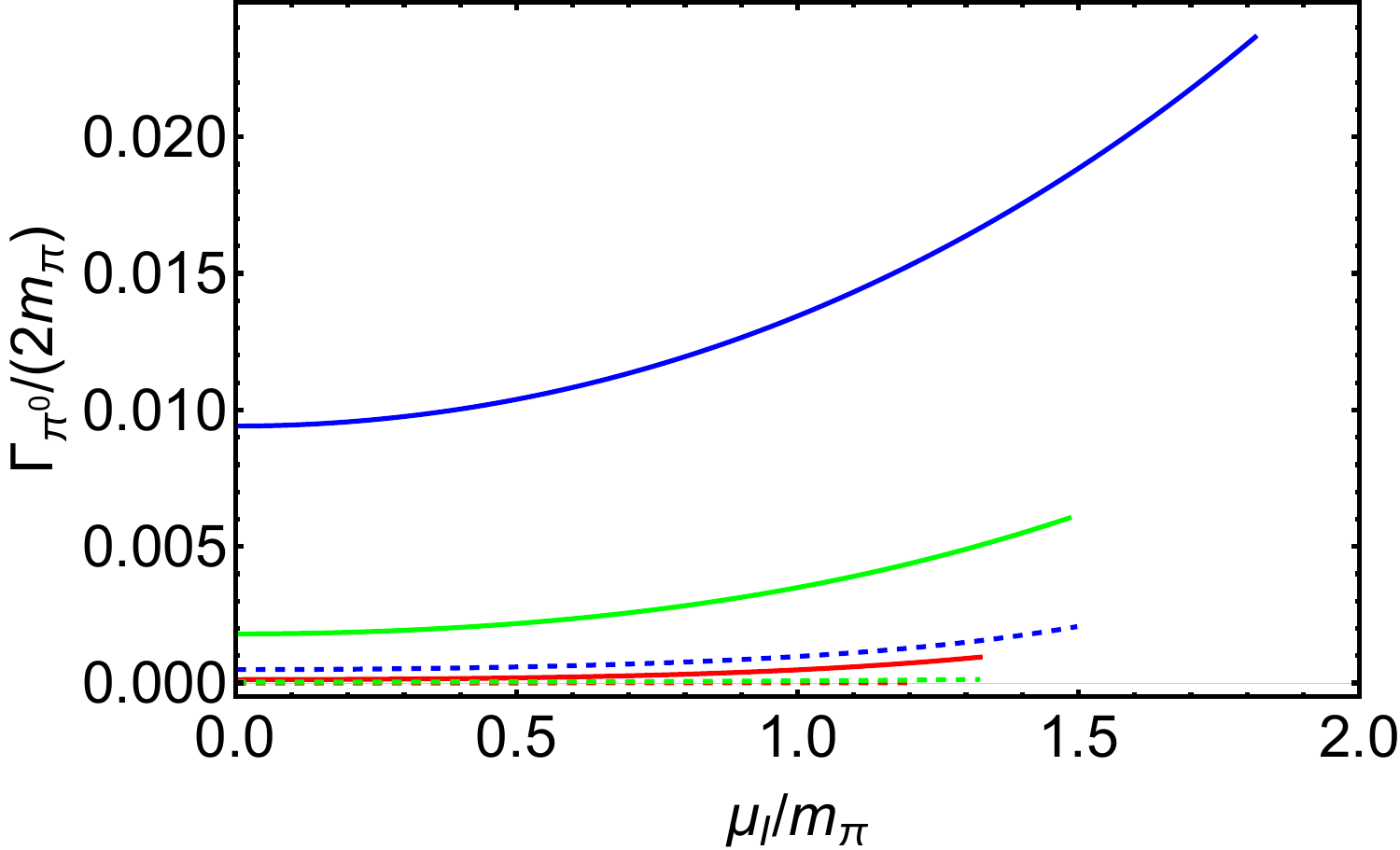}
        \put(88,55){\bf{(b)}} 
    \end{overpic}
    \begin{overpic}[width=0.40\textwidth]{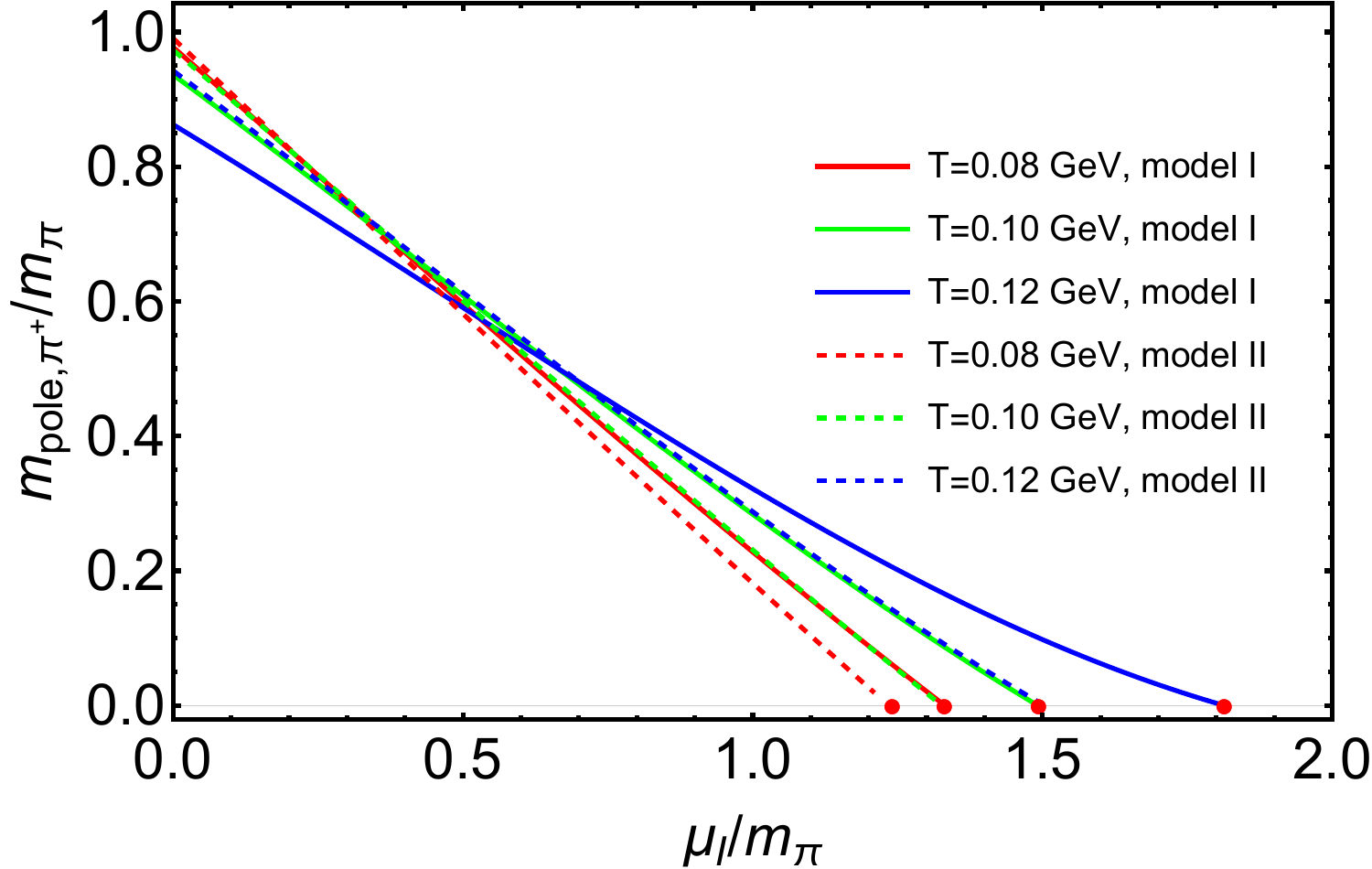}
        \put(88,55){\bf{(c)}} 
    \end{overpic}
     \begin{overpic}[width=0.42\textwidth]{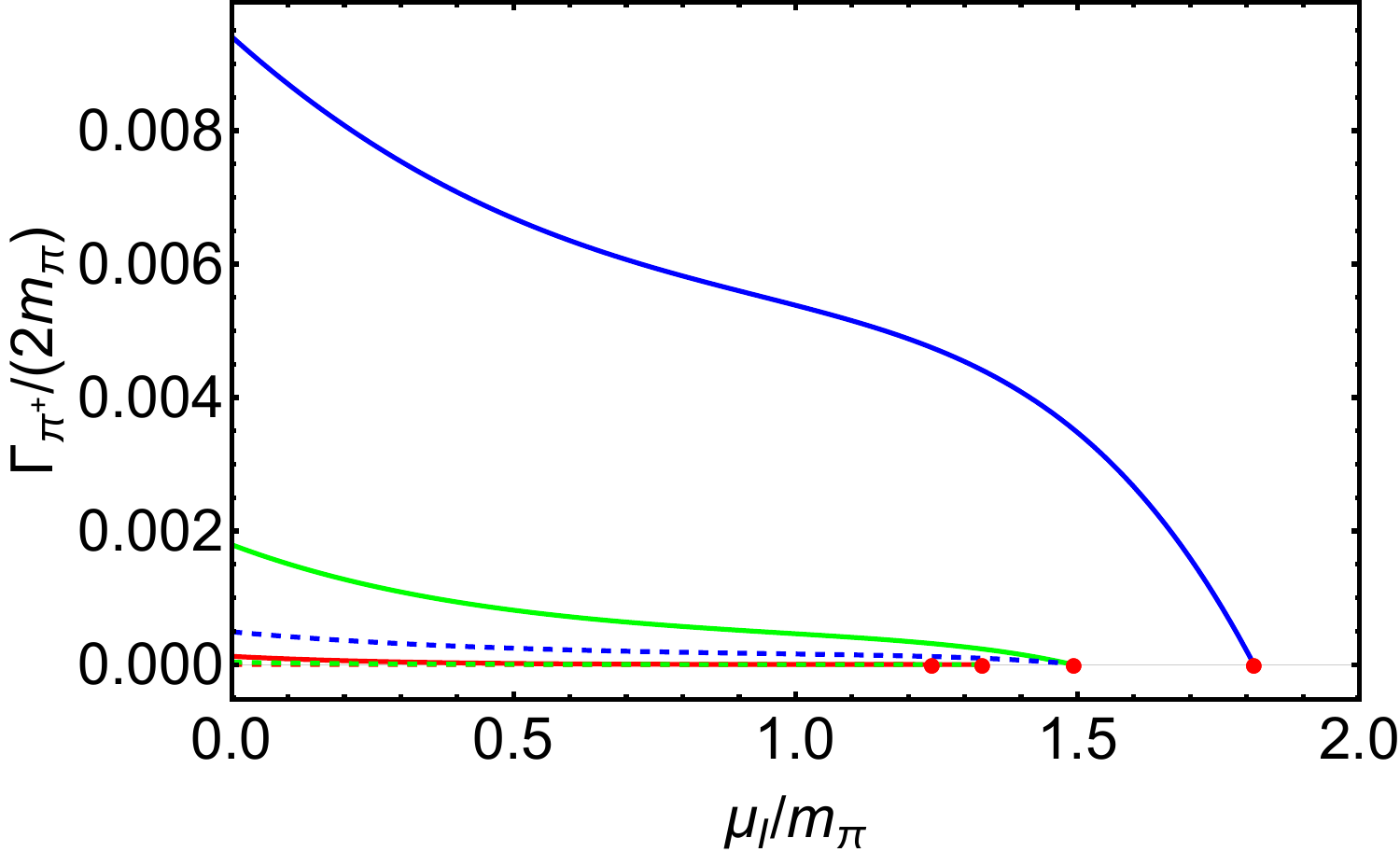}
        \put(88,55){\bf{(d)}} 
    \end{overpic}
     \begin{overpic}[width=0.40\textwidth]{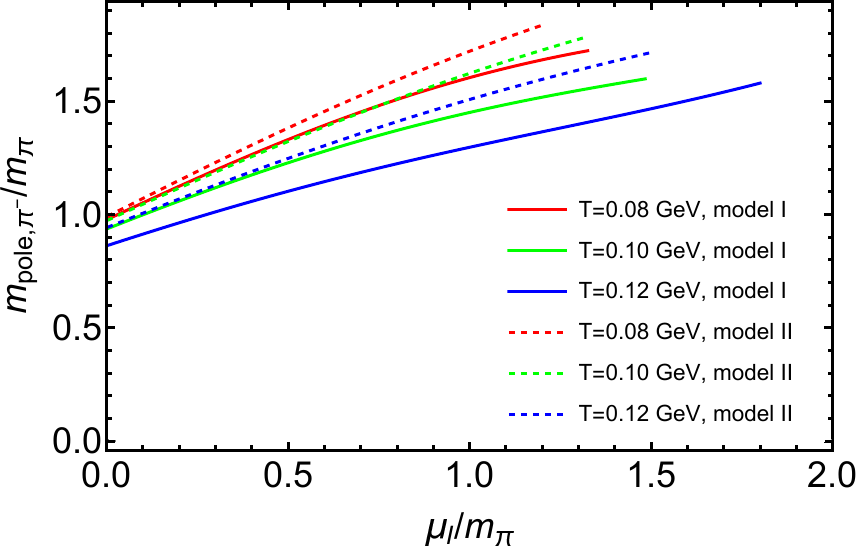}
        \put(88,55){\bf{(e)}} 
    \end{overpic}
     \begin{overpic}[width=0.42\textwidth]{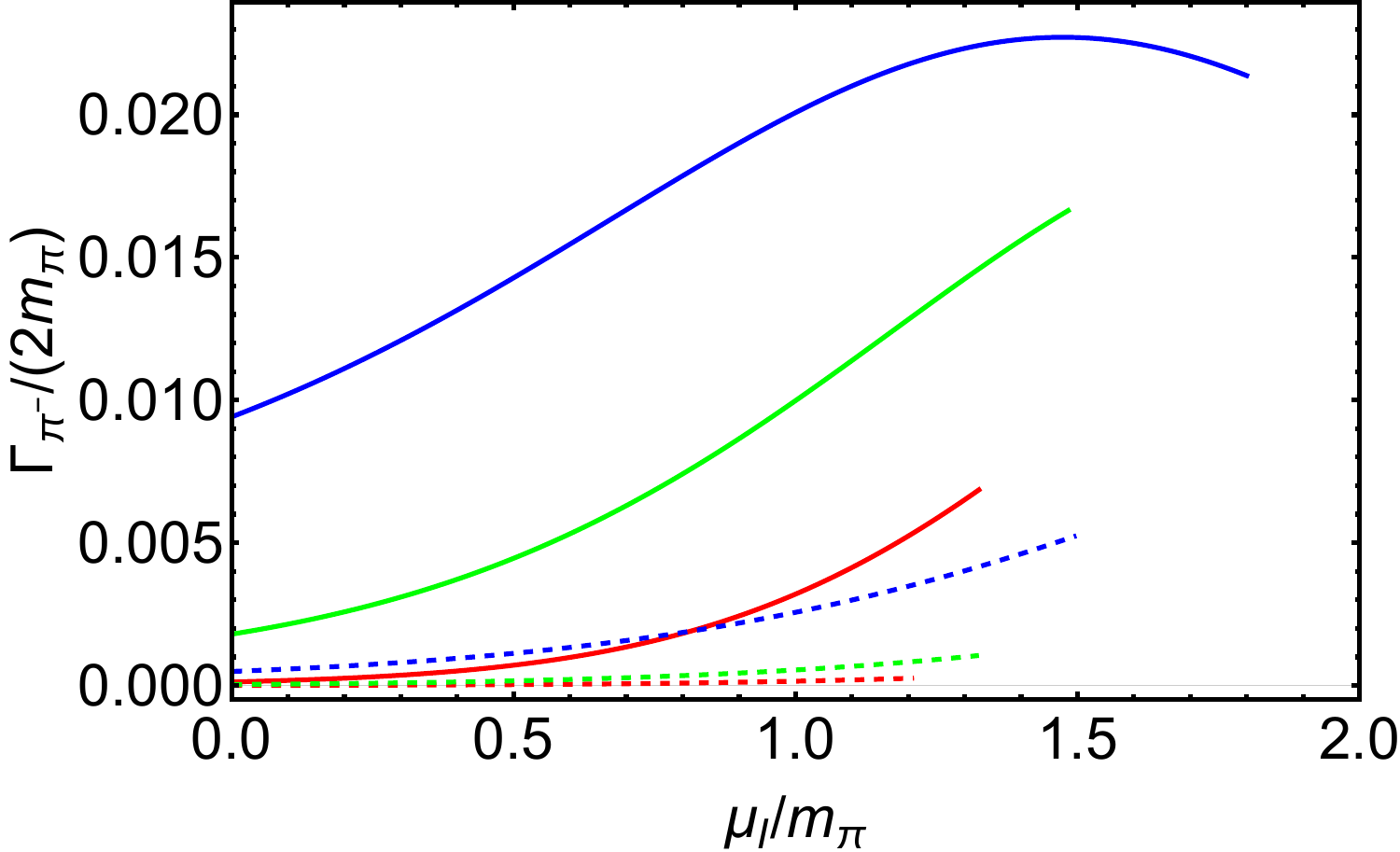}
        \put(88,55){\bf{(f)}} 
    \end{overpic}
    \caption{Pole mass and thermal width as functions of $\mu_{I}$ for \textbf{(a)},\textbf{(b)} $\pi^0$; \textbf{(c)},\textbf{(d)} $\pi^+$ and \textbf{(e)},\textbf{(f)} $\pi^-$. The red, green and blue lines stand for results at $T=0.08$, $0.10$, $0.12$ GeV, respectively. The solid lines represent results in model I while the dashed lines represent results in model II. $m_{\rm{pole},\pi^+}$ and -$\Gamma_{\pi^+}/2$ vanish at $\mu_{I}/m_{\pi}=1.32$, 1.49 and 1.82, respectively, in model I, and at $\mu_{I}/m_{\pi}=1.24$, 1.31 and 1.49, respectively, in model II, as shown by the red dots. 
    }
    \label{fig7}
\end{figure*}

\begin{figure}[htbp]
\centering
 \begin{overpic}[width=0.42\textwidth]{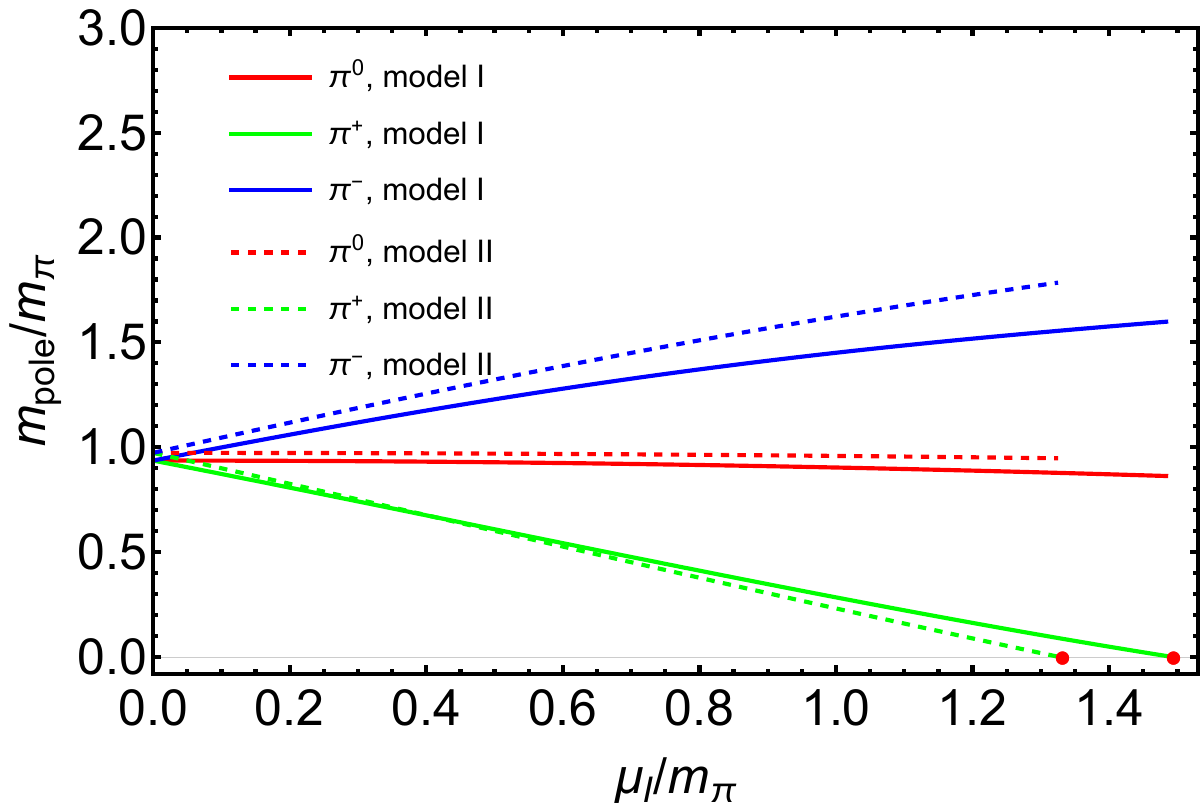} 
    \end{overpic}
\caption{\label{fig8} Pole masses as function of $\mu_{I}$ of the three modes $\pi^0, \pi^+, \pi^-$ at $T=0.10$ GeV in model I (represented by solid lines) and model II (represented by dashed lines). $m_{\rm{pole}}$ vanishes at $\mu_{I}^c=1.49 m_{\pi}$ for model I and 1.32 $m_{\pi}$ for model II, as shown by the red dots.}
\end{figure}

\subsubsection{The isospin density effect}

In this subsection, we would turn to the isospin chemical potential effects on pole masses and thermal width of pions. As shown in Figs.~\ref{fig7}(a) and~(b), both models possess the same trend for $\pi^0$. $m_{\rm{pole},\pi^0}$ and $\Gamma_{\pi^0}$ are monotonously varying with the increasing of $\mu_{I}$.  We find that  $\mu_{I}$ have slight effects on $m_{\rm{pole},\pi^0}$ and $\Gamma_{\pi^0}$ at low temperature, which are consistent with the analysis in Ref.~\cite{Cao:2020ryx}~\footnote{In which the effective quasiparticle masses of pions are extracted from the spectral functions.}. This result may be reasonable since $\pi^0$ has no isospin charge and is almost independent on $\mu_{I}$. The slight contribution for the isospin effect comes from the gravity background metric.

As for positive charged pions $\pi^+$, the numerical results are displayed in Figs.~\ref{fig7}(c) and~(d). We find that $m_{\rm{pole},\pi^+}$ decreases to zero almost linearly with the increasing $\mu_{I}$. The thermal width $\Gamma_{\pi^+}$ also decreases to zero with the increasing $\mu_{I}$. In model I, as shown in Fig.~\ref{fig7}(c) , both $m_{\rm{pole},\pi^+}$ and $\Gamma_{\pi^+}$ decrease to zero when $\mu_{I}$ increase to $\mu_{I}^{c}=0.184$ GeV with fixed $T=0.08$ GeV. This reflects the instability of the system and the happening of pion superfluid phase transition. As the matter of fact, the points at which $\pi^+$ becomes massless boson is exactly on the boundary between the pion condensed phase and normal phase. In addition, we have the  critical isospin chemical potential as
\begin{eqnarray}
T/ {(\rm GeV)} &=&0.08\quad \quad 0.10\quad\quad {\rm and}\quad\quad 0.12 , \nonumber\\
{\rm  Model\  I:}\ \mu_I^c/m_\pi &=&1.32 \quad \quad1.49 \quad \quad{\rm and }\quad\quad 1.82,\nonumber\\
{\rm  Model\  II:} \mu_I^c/m_\pi &=&1.24 \quad \quad 1.31 \quad \quad{\rm and }\quad\quad1.49.\nonumber
\end{eqnarray}
Above $\mu_{I}^c$, the $\rm{U}_{I}(1)$ symmetry, which is a subgroup of isospin $\rm{SU}(2)_I$, is broken and leads to the massless Goldstone boson $m_{\rm{pole},\pi^+}=0$. This result is consistent with the NJL model~\cite{Jiang:2011aw}. However, since we did not take back reaction of the pion condensate into account, the study of the masses at $\mu_{I}>\mu_{I}^c$ will be left for our future work.

The numerical results for the negative charged pions $\pi^-$ are shown in Figs.~\ref{fig7}(e) and (f). We find that $m_{\rm{pole},\pi^-}$ increases monotonously with the increasing of $\mu_{I}$ at fixed temperatures. The thermal width $\Gamma_{\pi^-}$ also increase monotonously with the increasing of $\mu_I$ for model I and for model II at relatively low temperatures. When the fixed temperature is high, e.g. $T=0.12$ GeV in model II, the thermal width first increases with the increasing of $\mu_I$ and then turn to decreases


Finally, we show the $\mu_{I}$ dependence of the three modes' pole masses together in Fig.~\ref{fig8} at fixed temperature $T=0.10$ GeV in both soft-wall AdS/QCD models. The qualitative behaviors are all the same in both soft-wall AdS/QCD models. As a result of the explicit ${\rm SU}_I(2)$ symmetry breaking, the pole mass splitting at finite isospin chemical potential, which is consistent with the previous study in the hard-wall model~\cite{Lee:2013oya} and the NJL model~\cite{Xia:2013caa}, as well as the quasi-particle masses extracted from the peak of the spectral function in the soft-wall model~\cite{Cao:2020ryx}. The pole mass of $\pi^+$, which is the goldstone boson of the pion superfluid phase transition, goes to zero at $\mu_I^c$. $\pi^0$ almost keeps invariant. At last, the pole mass of $\pi^-$ increases with the increasing $\mu_I$.

\section{\label{sec4}conclusion}
In this work, based on the soft-wall AdS/QCD models, we investigate the temperature and isospin chemical potential dependence of the pion quasipariticle masses (screening mass, pole mass and thermal width) in the chiral breaking phase ($T<T_{cp}$) and normal phase ($\mu_{I}<m_{\pi}$). Furthermore, we investigate the relationships between the Goldstone boson and the pion superfluid phase transition. A comparative study on the two kinds of models have been shown. Both models can provide experimental data at zero temperature. The two models give consistent results. Therefore, these may be qualitative behaviors shared by all the soft-wall AdS/QCD models.

On the one hand, we study the temperature dependence of the srceening mass at fixed isospin chemical potentials. The results show that $m_{\rm{scr},\pi^0}$ and $m_{\rm{scr},\pi^{\pm}}$ will split at finite $\mu_{I}$, but $m_{\rm{scr},\pi^{\pm}}$ are degenerate in the normal phase. In this case, the screening masses of charged pions are lower than the neutral pion at the same temperature. Both $m_{\rm{scr},\pi^0}$ and $m_{\rm{scr},\pi^{\pm}}$ increase monotonously with the increasing of the temperature, and the difference between them decreases when $T$ gets closed to $T_{cp}$. Since $\pi^0$ carries no isospin charge, $\mu_{I}$ has little impact on $m_{\rm{scr},\pi^0}$ which keeps unchanged with the increasing of $\mu_{I}$. While $m_{\rm{scr},\pi^{\pm}}$ decrease with the increased $\mu_{I}$, and reach zero at critical isospin chemical potential $\mu_{I}^c$. This implies a instability and the occurrence of pion superfluid phase transition. On the boundary between the normal phase and the pion superfluid phase, the ${\rm{U}_{I}}(1)$ symmetry is spontaneously broken, which leads to the appearance of the massless Goldstone boson $\pi^+$.

On the other hand, we also investigate the pole masses and thermal widths, at finite temperature and isospin chemical potential, extracting from the corresponding QNMs. The results suggest that the pole masses and thermal widths of $\pi^0, \pi^+, \pi^-$ will split at finite $\mu_{I}$. We found that $m_{\rm{pole},\pi^0}$ depends very weakly on $\mu_{I}$, since it carries no isospin charge.$m_{\rm{pole},\pi^+}$ decreases with the increasing of $\mu_{I}$ almost linearly and vanishes at the critical chemical potential $\mu_{I}^c$, where $\pi^+$ becomes a massless Goldstone boson. This implies the occurrence of pion superfluid phase transition.  At low temperature phase, $m_{\rm{pole},\pi^-}$ increases monotonously with the increasing of $\mu_{I}$, almost linearly. While $T$ turns high, $m_{\rm{pole},\pi^-}$ first increases and then decreases with the rise of $\mu_{I}$.  Both $m_{\rm{pole},\pi^0}$ and $m_{\rm{pole},\pi^-}$ decrease monotonously with the increasing of $T$. As for $\pi^+$, when $\mu_{I}$ is small, $m_{\rm{pole},\pi^+}$ also decreases monotonously with the increasing of $T$. However, when $\mu_{I}$ gets larger,  $m_{\rm{pole},\pi^+}$ first increases to a certain maximum and then decreases with the rise of $T$.  The thermal widths of the three modes increase with temperature.
However, we do not consider the pionic masses in the high-temperature phase as well as in the pion superfluid phase, which we leave this to our future work.

\begin{acknowledgments}
X.C. is supported by the National Natural Science Foundation of China under Grant Nos. 12275108 and the Fundamental Research Funds for the Central Universities under Grant No. 21622324. H.L. is supported by the National Natural Science Foundation of China under Grant No. 11405074. D.L. is supported by the National Natural Science Foundation of China under Grant Nos. 12275108, 12235016, 11805084.
\end{acknowledgments}

\bibliography{2refs}

\end{document}